\documentclass[sigconf]{acmart}

\usepackage{filecontents}

\usepackage{paralist}

\usepackage{hyperref}

\usepackage{tabularx}
\usepackage{mathrsfs}
\usepackage{tikz}
\usepackage{amsmath}
\usepackage{comment}

\usepackage{enumitem}
\usepackage{algorithm}
\usepackage{algpseudocode}
\usepackage{float}

\usepackage{caption}
\usepackage{subcaption}

\usepackage{filecontents}
\usepackage{siunitx}
\usepackage{varwidth}
\usepackage{amsmath}
\usepackage{multirow}
\usepackage{framed}
\usepackage{mdframed}
\usepackage{mathtools, nccmath}
\usepackage{balance}
\usepackage{threeparttable}
\usepackage{amsfonts}
\usepackage{mathrsfs}
\usepackage{tikz}
\usepackage{float}
\usepackage{amsmath}
\usepackage{comment}  
\usepackage{algpseudocode}
\usepackage{float}
\usepackage{filecontents}
\usepackage{amsmath} 
\usepackage{multirow}
\usepackage{framed}
 
\usepackage{picture,xcolor}

\usepackage[export]{adjustbox}

\usepackage{mathtools, nccmath}
\usepackage{xspace}
\usepackage{booktabs,siunitx,caption}
\usepackage{varwidth}
\usepackage{placeins}
\usepackage{algpseudocode}
\long\def\authornote#1{%
  \leavevmode\unskip\raisebox{-3.5pt}{\rlap{$\scriptstyle\diamond$}}%
  \marginpar{\raggedright\hbadness=10000
    \def\baselinestretch{0.8}\tiny
    \it #1\par}}

\usepackage{breakurl}

\usepackage{tabularx}
\makeatletter
\renewcommand{\ALG@name}{Algorithm}
\makeatother

\newcommand{\sysname}{\emph{MembershipTracker}\xspace}

\usepackage{tikz}
\usepackage{ifluatex,ifxetex}
\ifluatex
  \usepackage{fontspec} 
\else\ifxetex
  \usepackage{fontspec} 
\else 
  \usepackage[T1]{fontenc}
  \usepackage[utf8]{inputenc} 
\fi\fi

\usepackage{pifont}
\newcommand{\cmark}{\ding{51}}%
\newcommand{\xmark}{\ding{55}}%
\usepackage{tikz}

\usepackage{multicol}
\usepackage{wrapfig}

\usepackage{tikz}
\newcommand*{\circled}[2][]{\tikz[baseline=(C.base)]{
    \node[inner sep=0pt] (C) {\vphantom{1g}#2};
    \node[draw, circle, inner sep=1pt, yshift=1pt] 
        at (C.center) {\vphantom{1g}};}}

\algnewcommand\algorithmicforeach{\textbf{for each}}
\algdef{S}[FOR]{ForEach}[1]{\algorithmicforeach\ #1\ \algorithmicdo}
\algnewcommand{\IIf}[1]{\State\algorithmicif\ #1\ \algorithmicthen}
\algnewcommand{\EndIIf}{\unskip\ \algorithmicend\ \algorithmicif}

\copyrightyear{2025}
\acmYear{2025}
\setcopyright{acmlicensed}\acmConference[CCS'25]{ACM Conference on Computer and Communications Security}{October 13--17, 2025}{Taipei, Taiwan}
\acmBooktitle{ACM Conference on Computer and Communications Security (CCS'25), October 13--17, 2025, Taipei, Taiwan}

\begin{document}


\title{{Anonymity Unveiled: A Practical Framework for Auditing Data Use in Deep Learning Models}}

\author{Zitao Chen} 
 
  \affiliation{\institution{University of British Columbia}
      \country{Vancouver, BC, Canada}} 
\email{zitaoc@ece.ubc.ca}

\author{Karthik Pattabiraman}
\affiliation{\institution{University of British Columbia}
      \country{Vancouver, BC, Canada}} 
\email{karthikp@ece.ubc.ca}


\begin{abstract} 
The rise of deep learning (DL) has led to a surging demand for training data, which incentivizes the creators of DL models to trawl through the Internet for training materials. 
Meanwhile, users often have limited control over whether their data (e.g., facial images) are used to train DL models without their consent, which has engendered pressing concerns. 

This work proposes \sysname, a practical data auditing tool that can empower ordinary users to reliably detect the unauthorized use of their data in training DL models. 
We view data auditing through the lens of {membership inference} (MI). 
\sysname consists of a lightweight data marking component to mark the target data with small and targeted changes, which can be strongly memorized by the model trained on them; and a specialized MI-based verification process to audit whether the model exhibits  strong memorization on the target samples.

\sysname only requires the users to mark a small fraction of data (0.005\%$\sim$0.1\% in proportion to the  training set), and it enables the users to {reliably} detect the unauthorized use of their data (average 100\% TPR@0\% FPR). We show that \sysname is highly effective across various settings, including industry-scale training on the full-size ImageNet-1k dataset. We finally evaluate \sysname under multiple classes of countermeasures\footnote{Our code is available at \url{https://github.com/DependableSystemsLab/MembershipTracker}.}.
\end{abstract}


\maketitle

\section{Introduction}
\label{sec:intro}

Modern deep learning (DL) models have attained remarkable performance in various tasks such as image classification and language generation.  
Their success is largely driven by the availability of massive training data~\cite{rombach2022high,ramesh2022hierarchical}. 
However, it is not always clear whether the data used to train these models has been used with the permission of the data owners. 
For instance, Clearview.ai, in developing their facial recognition system, scraped billions of user photos from social media platform without the users' consent~\cite{clearviewai}. 
The unregulated use of personal data for building DL models has engendered mounting concerns~\cite{ai-art-lawsuit,midjourney,clearviewai}, and is in violation of privacy laws such as the European Union's General Data Protection Regulation (GDPR)~\cite{gdpr}. 
In spite of this, data holders often have limited agency in {detecting} the unauthorized use of their data {due to the lack of practical data provenance tools}.

This leads to two main lines of work for data provenance. 
The first line of work seek to inject designated features into the user data, which can induce the model to exhibit a certain detectable behavior for provenance purposes~\cite{sablayrolles2020radioactive,hu2022membership,li2022untargeted,wenger2022data}.  
The injected features can take the form of a backdoor trigger pattern
~\cite{hu2022membership,li2022untargeted}; or a spurious feature pattern that can cause output probability shift by the model~\cite{wenger2022data}. 
However, these techniques either impose strong assumptions on the users (e.g., the ability to mark a large fraction of data)~\cite{sablayrolles2020radioactive,li2022untargeted,hu2022membership} and/or {suffer from limited auditing performance~\cite{sablayrolles2020radioactive,li2022untargeted,wenger2022data}.

The second line of work is based on \emph{membership inference} (MI). 
{There are  instance- and user-level MI - the former infers whether a \emph{specific} sample is used to train a model~\cite{shokri2017membership,carlini2022membership}; while the latter detects the usage of \emph{any} of the user's data~\cite{song2019auditing,chen2023face,kandpal2023user}. 
Our work focuses on auditing whether \emph{specific user samples} are used for training the target model. 
}
Unfortunately, existing MI methods often suffer from limited effectiveness, as measured by their true positive rate (TPR) and false positive rate (FPR)~\cite{carlini2022membership}. However, a high-power MI method is needed to facilitate  data provenance with {low FPR} (to avoid false accusation) \emph{and} high TPR (for correct data provenance). 

To improve, recent studies propose several techniques based on  data poisoning~\cite{tramer2022truth,chen2022amplifying,chaudhari2023chameleon}. 
Their idea is to inject {mislabeled} samples into the model's training set, which can amplify the  model's behavioral changes on some target samples, thereby making them easier to be de-identified. 
To apply these techniques for data auditing, users {have to} be able to mislabel some training data, which, unfortunately, excludes the real-world scenarios where the data labels are assigned by the model owner or an external service~\cite{data-annotation-1,deng2009imagenet}. This renders these techniques challenging for real-world use.

\begin{figure}[t]
  \centering
  \includegraphics[  height=.9in]{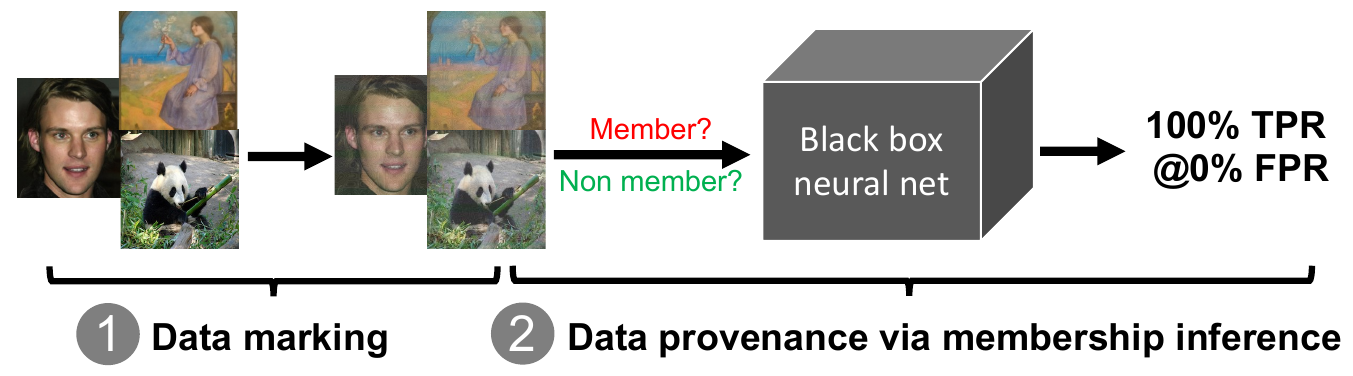}
  \caption{\sysname is a data provenance tool that operates by: (1) marking the target data   (e.g., facial images, artworks) with small and targeted changes, and then (2) initiating a specialized membership inference process to audit whether the target data are used for training the model. 
  }
  \label{fig:intro} 
  \vspace{-2mm}
\end{figure}

\textbf{This work}. We present \sysname, a MI-based data provenance tool that can overcome the aforementioned challenges, with a two-step process as illustrated in Fig.~\ref{fig:intro}.

\sysname operates under the \textbf{realistic setting} where we assume the users have {neither} the capability nor expertise to train shadow/proxy models~\cite{ye2021enhanced,bertran2024scalable,carlini2022membership},  
or to manipulate the data labels~\cite{tramer2022truth}, as these are well beyond the capacity of ordinary personal users (Section~\ref{sec:capability}). 
Rather, \sysname only requires the users to mark a small fraction of data (amounting to 0.005\%$\sim$0.1\% of the training set) for provenance purposes. 
Our goal is to empower the users to carry out high-power MI on their target data with high TPR and low FPR, given \emph{black-box} access to the model.

\textbf{Technical design.} 
We first observe that the effectiveness of MI on a given model is directly related to the model's ability in memorizing individual data points, and hence samples that are strongly memorized by the model (e.g., outlier samples) are easier to be de-identified~\cite{yeom2018privacy,carlini2022membership,tramer2022truth,chen2024method}. 
Therefore, in order to fulfill our goal, there are two challenges as follows. 

\begin{enumerate}[leftmargin=*]
\itemsep0em  
\item How can the users \emph{amplify} the model's memorization on the target data with minimal data modification? 
\item How can the users reliably \emph{audit} whether the model exhibits strong memorization on their target data?
\end{enumerate}

For the first challenge, we propose a lightweight data marking technique to mark the target data with small and targeted changes, which are formulated as the combination of outlier feature and procedural noise~\cite{co2019procedural}. These subtle changes can be readily created by the users without assuming access to a proxy model, and can still induce the model to strongly memorize the specially-marked samples while preserving their visual information (e.g., Fig.~\ref{fig:intro}).

{For the second challenge}, given a target user's small set of marked samples, we propose a novel \emph{set}-based MI process to audit whether they are used to train the model. 
{Our approach follows a common idea in existing user-level MI studies~\cite{song2019auditing,meeus2023did,kandpal2023user} to  leverage the \emph{collective} information across the small set of target samples for MI. But unlike prior methods, \sysname does not require any additional shadow/reference models~\cite{song2019auditing,meeus2023did,kandpal2023user} and can enable high-power MI for reliable data provenance. 
}

\textbf{Contributions}. We make three contributions as follows. 

\begin{itemize}[leftmargin=*]
\itemsep0em  
\item Propose a lightweight data marking technique to mark the users' target data with small and targeted changes, which can be strongly memorized by the model trained on them. 
\item Develop a high-power set-based MI verification process that can reliably audit whether the target data are used to train the model (with high TPR when controlled under a low FPR). 
\item Integrate the above as a tool called \sysname, and evaluate its effectiveness across a variety of settings (six benchmark datasets and {six DL architectures}).  
We also comprehensively evaluate a total of six classes of countermeasures. 
\end{itemize}

We find that by merely marking a small fraction of samples (0.005\%$\sim$0.1\% of the dataset), \sysname effectively empowers the users to reliably audit the usage of their data (average 100\% TPR@0\% FPR), and it is scalable to the large-scale ImageNet training (on both Convolutional and Transformer network). 
\emph{To the best of our knowledge, \sysname is the first technique that can scale to support reliable data auditing in large-scale training setting.} 
This renders \sysname a practical data auditing tool that contributes to responsible AI practice.

\section{Related Work}
\label{sec:related-work}
Our work focuses on detecting the unauthorized use of personal data in training DL models, which is complementary to existing work that aim to protect the data from unwanted use by rendering them un-learnable~\cite{shan2020fawkes,huang2021unlearnable}. 
Hence we focus on existing literature for data provenance, and we classify them into two categories.

\textbf{Non membership-inference based solutions.}
We divide them into dataset- and user-level solutions. 

\emph{Dataset level provenance} aims to detect whether a DL model is trained on a specific dataset~\cite{maini2020dataset,sablayrolles2020radioactive,li2022untargeted}. 
Maini et al. propose dataset inference, which detects whether two models trained on the same dataset exhibit similarities in their decision boundaries~\cite{maini2020dataset}. 
Other work modify portions of training set to leave a certain artifact on the models, such as radioactive data~\cite{sablayrolles2020radioactive} and backdoor watermark~\cite{li2022untargeted}. 
However, recent work~\cite{huang2024general} finds that these techniques~\cite{sablayrolles2020radioactive,li2022untargeted} still have poor auditing performance.  
Moreover, dataset level solutions require several assumptions that are outside the scope of our work. 
First, many of them require control over a nontrivial portion of training set (e.g., 10\%$\sim$20\%~\cite{sablayrolles2020radioactive,li2022untargeted}). 
Some solutions assume access to some known feature extractor~\cite{sablayrolles2020radioactive}, access to the training set or the ability to train a proxy model~\cite{maini2020dataset,sablayrolles2020radioactive}. 
These are well beyond the user capability we consider (in Section~\ref{sec:capability}).

\emph{User level provenance} seeks to detect whether the users' data are used to train a DL model~\cite{hu2022membership,wenger2022data}, and they require only modifying the users' own data. 
Their idea is to mark the users' data with a designated feature to induce a detectable behavior into the model. 
Hu et al. propose to inject backdoor trigger feature into the user data~\cite{hu2022membership}, but they assume the user capability to manipulate the data labels. Other studies on clean-label backdoor attacks may be repurposed to overcome this, but they still require training a proxy model to compute the poisoned data~\cite{souri2022sleeper,zeng2023narcissus}.

Wenger et al. propose to inject a target spurious feature into the user data to be learned by the suspected model, and then detect whether the model has associated the injected feature with the target class label~\cite{wenger2022data}. 
To detect, the users can separately overlay the target and non-target features to some auxiliary data (outside the target class), and then determine whether the target feature causes a slightly higher prediction probability on the target class, compared with the non-target features  (e.g., 10\% vs. 1\%). 
Both their work and ours involve adding spurious features to mark users' data, but there are several major differences between the two. 
Notably, our analysis reveals that they  \cite{wenger2022data} significantly \emph{underestimate} their detection false positives (from 0\% to >30\%). 
This is because they employ a \emph{discrepant} detection procedure for the testing features that are used for training (true positives), and those that are not (false positives); however, a fair evaluation should follow a consistent procedure (see our detailed analysis in Appendix~\ref{sec:wenger-comp}).

\textbf{Membership-inference (MI)  based solutions.} 
MI also represents a natural fit for tracing data provenance, and 
there are instance- and user-level MI solutions. 

\emph{Instance-level MI} detects whether a specific instance is used to train a model. It was first proposed by Shokri et al.~\cite{shokri2017membership} and there are many follow-up studies~\cite{yeom2018privacy,choquette2021label,bertran2024scalable,ye2021enhanced,carlini2022membership}. 
However, even state-of-the-art MI methods~\cite{carlini2022membership,bertran2024scalable,ye2021enhanced} still struggle with achieving high TPR when controlled at the low FPR regime (more in Section~\ref{sec:motivation}). 

This has led to several attempts at improving the effectiveness of existing MI methods via data poisoning. 
Tramer et al. propose to inject mislabeled samples into the model's training set, which can transform the target samples into outliers and amplify their influence on the model's decision, thereby making the target samples easier to be de-identified~\cite{tramer2022truth}. 
While injecting mislabeled samples represents an effective solution for improving the MI success~\cite{tramer2022truth,chen2022amplifying,chaudhari2023chameleon}, it may be challenging for the users to apply to their data, because in many scenarios the users may not have control over the data labels~\cite{data-annotation-1,wenger2022data,deng2009imagenet}. 
{Even if} this is feasible, we find that such a method can still only increase the MI success to a limited extent (validated in Appendix~\ref{sec:tramer-et-al-comparison}).  
By contrast, \sysname does not assume control of the data labels, and is still able to facilitate high-power MI for tracing data provenance.

\emph{User-level MI} aims to detect whether any of a user's data is used to train a model. 
Unlike instance-level MI, the auditor possesses a set of samples from a target user that are \emph{not} necessarily used to train the target model. 
Song et al.~\cite{song2019auditing} propose the first user-level MI for natural language models, which first trains multiple shadow models, and then uses the outputs from shadow models as the features to train an audit model. 
Subsequent work develop solutions for other settings, including speech recognition models~\cite{miao2021audio,chen2023slmia}, metric learning models~\cite{li2022user,chen2023face} and large language models~\cite{kandpal2023user,meeus2023did}. 

{
A common thread of existing user-level MI methods is to aggregate the information across multiple samples (e.g., different voices of the same speaker) to perform MI~\cite{song2019auditing,chen2023slmia,chen2023face,kandpal2023user}. 
The MI process in \sysname follows a similar philosophy. 
However, prior work often requires access to additional reference models~\cite{kandpal2023user} or shadow models~\cite{song2019auditing,chen2023face,chen2023slmia}, and they still have limited MI effectiveness (e.g., limited TPR under the low FPR regime~\cite{kandpal2023user,meeus2023did}). 
\sysname overcomes these limitations by means of a novel data-marking and specialized MI auditing process, which can simultaneously achieve  high TPR and low FPR for data provenance. 
}

Recent work by Huang et al.~\cite{huang2024general} proposes a general framework for auditing unauthorized data use in training DL models. Their idea is to generate two perturbed copies of the target data, randomly publish one of them, and then  compare the model's membership score on the published vs. the unpublished one. 
{
While their framework can work for different domains such as foundation models, they still require the data holders to mark a substantial portion (1\%$\sim$10\%) of dataset. 
Instead, \sysname considers the more challenging (and realistic) setting where the marked data amounts to $\leq0.1\%$ of the  dataset. We also ``stress test'' their technique under a similar setting as ours and find that their performance degrades considerably then (see Appendix~\ref{sec:huang-et-al}). 
}

\section{Problem Formulation}
\label{sec:background}

\emph{Membership Inference (MI) Game}. 
We denote ${F}_{\theta} : \mathcal{X}\rightarrow[0,1]^n$ as a model that maps an input sample $x \in\mathcal{X}$ to a probability vector over $n$ classes. 
$D=\{(x_i,y_i)\}_{i=1}^n$ is a training set sampled from some distribution $\mathcal{D}$. 
${F}_{\theta}\leftarrow \mathcal{T}(D)$ denotes a  model ${F}_{\theta}$ produced from running the training algorithm $\mathcal{T}$ on $D$. 

Next, we define a MI game, which proceeds between a challenger $\mathcal{C}$ and an adversary $\mathcal{A}$. 
{$\mathcal{A}$ has to guess which element from some universe $\mathcal{U}$ was used to train a model. For a specific target point $x$, $\mathcal{U} = \{x, \perp\}$, where $ \perp$ indicates the absence of an example}.

\begin{enumerate}
  \itemsep0em 
\item The challenger samples a dataset $D\leftarrow\mathcal{D}$, and a target point $z\leftarrow \mathcal{U}$ (such that $D \cap \mathcal{U} = \emptyset$).
\item The challenger trains a model ${F}_{\theta}\leftarrow \mathcal{T}(D\cup \{z\})$ on the dataset $D$ and target point $z$. 
\item The challenger gives the adversary query access to ${F}_{\theta}$.
\item The adversary emits a guess $\hat{z}\in \mathcal{U}$.
\item The adversary wins the game if $\hat{z}=z$.
\end{enumerate}

\subsection{MI for Data Provenance} 
We now describe the MI game for data provenance, which proceeds between a challenger (model creator) $\mathcal{C}$ and a target user ${U}$. 
We consider image classification in this work, and ${U}$ seeks to audit whether his/her images (denoted as $X$) are used by $\mathcal{C}$ to train a model $F$ without permission. 
We add the ability for ${U}$ to modify $X$ for provenance purpose.

\begin{enumerate}
  \itemsep0em 
\item {The user is allowed to modify ${X} \in \mathcal{U}$} ($\mathcal{U}=\{ {X} , \perp\}$). 
\item The challenger samples a dataset $D\leftarrow\mathcal{D}$, and a target set $Z\leftarrow \mathcal{U}$ (such that $D \cap \mathcal{U} = \emptyset$). 
\item The challenger trains a model ${F}_{\theta}\leftarrow \mathcal{T}(D\cup \{Z\})$ on the dataset $D$ and target set $Z$. 
\item The challenger gives the user query access to ${F}_{\theta}$.
\item The user emits a guess $\hat{Z}\in \mathcal{U}$.
\item The user wins the game if $\hat{Z}=Z$.
\end{enumerate}

For a given user $U$, $\mathcal{U} = \{X, \perp\}$, where $ \perp$ indicates the absence of the target samples by $U$, and the  goal is to determine whether $F$ is trained on $D$ or $D\cup \{X\}$. 

{
Unless otherwise stated, we follow the common practice to measure the MI success by its true positive rate (TPR) controlled under a low false positive rate (FPR)~\cite{carlini2022membership,ye2021enhanced}.
} 

\subsubsection{\textbf{User capability}}
\label{sec:capability}

We consider ordinary personal users ${U}$ with limited ML expertise and resources, and they are willing to add visual distortion to their data for provenance purposes. 
We next specify their capabilities and constraints. 

First, we assume $U$ can only mark a small number of samples, and these are to be included as part of a larger dataset. 
This simulates the realistic scenario where the dataset is curated from multiple data sources (e.g., \cite{deng2009imagenet}), and thus data contributed by each user constitutes only a small portion of the dataset (e.g., 0.1\%). 

Next, in the provenance verification process, we assume $U$ has access to a set of non-member samples that are not used for training the target model\footnote{{Under the presumption that only a small fraction of data sampled from the distribution were used in training, one may simply draw a random sample from the underlying distribution, and be confident that it is representative of the non-member data~\cite{bertran2024scalable,ye2021enhanced}.}}, 
which is a common assumption\footnote{{While we also adopt  this common assumption in our study, we  conduct additional experiments and find that this assumption can in fact, be relaxed (see Section~\ref{sec:limitations}).}} to control the MI process at a low FPR regime. 
$U$ also has black-box access to query the target model. 
In a real world setting, users often have limited control and knowledge beyond their own data, and thus we specify the constraints by $U$ as follows. 

\begin{itemize}[leftmargin=*]
\itemsep0em  
\item $U$ cannot modify their data's labels, which are assigned by the model owner or an external service~\cite{data-annotation-1,wenger2022data,deng2009imagenet}. 
\item $U$ has no access to $D$ (i.e., the training data by other users).
\item $U$ has no knowledge of the black-box model $F$. 
\item $U$ has no expertise/resources to train any shadow/proxy models, as in many settings, training even a single shadow model can impose prohibitively high data and compute requirements (especially for the large models)~\cite{tramer2022truth}. 
\end{itemize}

\section{Methodology}
\label{sec:methodology}

We first conduct an empirical study to understand the limitations of leveraging existing MI methods for data auditing, and then describe our design goals. 
We then present \sysname with a data marking and provenance verification process: first by 
 marking the users' protected data with targeted changes (Section~\ref{sec:data-mod}); then performing a specialized MI process to audit whether the marked data are used to train the model (Section~\ref{sec:mi}).

\textbf{Motivation}. 
Our work addresses data provenance through MI, and there are several available MI methods~\cite{ye2021enhanced,bertran2024scalable,carlini2022membership,tramer2022truth}. 
However, these methods still suffer from limited MI effectiveness in the low FPR regime and/or impose strong assumptions on the users - we elaborate both limitations next.

\circled{1} Existing MI methods achieve limited TPR under a low FPR. 

To validate this, we use two methods in the state-of-the-art Likelihood-ratio attack, one is with and the other without shadow model calibration (128 shadow models)~\cite{carlini2022membership}. 
We evaluate both methods on a WideResNet-28-4 model trained on half of the training set in CIFAR100 dataset. 
As shown, even the shadow-model based method can only achieve 27.21\% TPR@0.1\% FPR.

This is undesirable because 
achiving high TPR (under a low FPR) is crucial for the users to reliably detect the unauthorized use of their data. 
As in Fig.~\ref{fig:lira}, existing MI methods struggle with this task. The reason is that the training samples are not always strongly memorized by the model (except some outliers)~\cite{yeom2018privacy,carlini2022membership,tramer2022truth}, which is directly related to the ability of performing accurate MI. 
This problem motivates related work~\cite{tramer2022truth,chen2022amplifying,chaudhari2023chameleon} to amplify the model's memorization on the training samples, and we discuss their limitation next. 

\begin{figure}[t]
  \centering
  \includegraphics[height=1.3in]{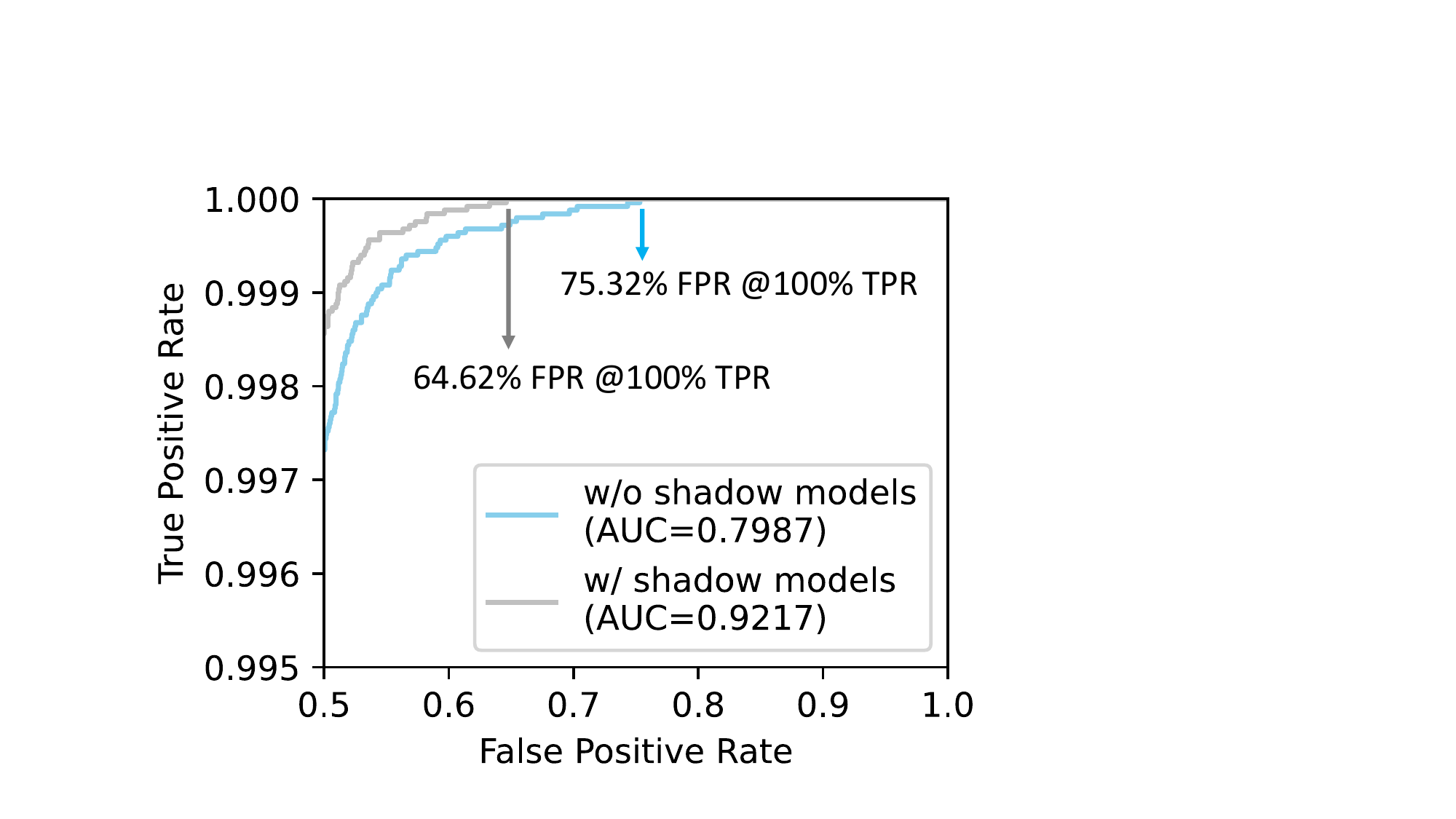}
  \caption{State-of-the-art MI methods~\cite{carlini2022membership} achieve limited TPR under the low FPR regime (undesirable). 
  }
  \label{fig:lira} 
  \vspace{-2mm}
\end{figure}

\circled{2} State-of-the-art MI methods either require training shadow models or assume control to manipulate the data labels, both are beyond ordinary users' capabilities. 
For example, leading MI methods  \cite{carlini2022membership,bertran2024scalable,ye2021enhanced} commonly require training shadow models to calibrate the inference process, which assumes additional data and compute access, and is challenging for the non-expert users.

There are other methods that do not require shadow-model calibration, but they either suffer from poor effectiveness~\cite{yeom2018privacy,carlini2022membership} or impose unrealistic assumption on the users~\cite{tramer2022truth,chaudhari2023chameleon}. 
E.g., the approach by Tramer et al.~\cite{tramer2022truth} can improve the MI success without shadow-model calibration, but it assumes users have the ability to mislabel some training data, which can be challenging in the cases where the users have no control over the data labels~\cite{data-annotation-1,wenger2022data,deng2009imagenet}. Even if this is feasible, we find that the increase of MI success is still largely limited (see Appendix~\ref{sec:tramer-et-al-comparison} for a validation).

\textbf{Design Goals}.
\label{sec:goal}
There are three design goals in our work. 

Goal (1): \emph{ High provenance effectiveness.} Our main goal is to enable those users whose data are misused for model training to perform successful data provenance (high TPR), and we also want to ensure low FPR to avoid false accusation. Therefore, our goal is to achieve high TPR@low FPR. 

Goal (2): \emph{ Preserving high visual quality.} The modification created to mark the users' data should not severely distort their visual information, e.g., the identity in a facial image should still be easily recognizable. 

Goal (3): \emph{ Minimal technical requirements.} The provenance tool should be usable by ordinary users even with limited expertise and resources (specified in Section~\ref{sec:capability}).

\subsection{Data Marking}
\label{sec:data-mod}
The goal of data marking is to ensure that the membership of the  marked samples can be reliably de-identified. 
To this end, we first observe that the ability to perform accurate MI is directly related to the model's propensity in memorizing individual data points, and data that are strongly memorized by the model (e.g., outlier samples, mislabeled samples) are known to be easier to be de-identified~\cite{carlini2022membership,yeom2018privacy,tramer2022truth,chen2024method}. 
Based on this, our goal can be formulated as: how can the users modify their data to amplify the model's memorization on the target data while preserving high visual quality?

In Section~\ref{sec:initial-approach}, we first present an initial approach that is highly effective in inducing the target samples to be memorized by the model, but it also completely destroys the visual information in the data. 
We use this effective yet unrealistic method as a strawman approach, and then introduce our proposal of a two-step solution, which can greatly reduce the visual distortion to the data while still fulfilling our goal.

\subsubsection{\textbf{An initial strawman approach}}
\label{sec:initial-approach}
Inspired by the observation that atypical samples presented in the dataset are easier to be de-identified (e.g., mislabeled samples~\cite{tramer2022truth}, or samples that are out-of-distribution - OOD~\cite{carlini2022membership}), a straightforward approach is to directly \emph{replace} the original target samples as OOD samples. 

There are various ways to generate OOD samples, and we develop a simple method that creates samples with random color stripes, such as the one shown below (the method's details are in Section~\ref{sec:exp-setup}). Meanwhile, other alternative methods such as using samples from an OOD dataset may also be considered (Appendix~\ref{sec:ablation-diff-ood}). 

\begin{wrapfigure}{R}{0.2\textwidth}
\centering
\includegraphics[width=0.2\textwidth]{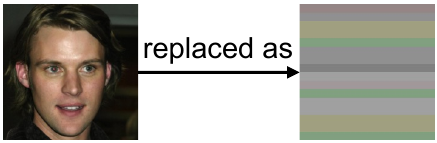}
\label{fig:initial}
\vspace{-2mm}
\end{wrapfigure}

We first conduct an experiment to validate the effectiveness of the above approach. 
We use the CIFAR100 dataset with half of the training set for model training. 
We consider 100 target users from different classes replacing their data as some random OOD samples, and each contributes a small number of samples to the training set (0.1\% in proportion). 
In terms of the signal function for MI, we follow prior work~\cite{sablayrolles2019white,ye2021enhanced} to use {prediction loss} values. 

To perform MI, we compare the prediction loss on the target OOD samples (from the {member} users) and some non-target OOD samples (from the {non-member} users). 
We find that the model indeed strongly memorizes those OOD samples present in the training set, for which we observe a 100\% TPR@1\% FPR. 
In comparison, if the target data are unmodified, the model only exhibits weak memorization on them, which yields merely 1.28\% TPR@1\% FPR. 

\emph{Limitation.} 
Despite its effectiveness, this approach completely removes the visual information in the data (e.g., the facial identity is no longer recognizable), and they are also easy to notice. 
Next, we present two methods that can greatly reduce the visual distortion to the target data, while still enabling them to be strongly memorized by the model (Fig.~\ref{fig:data-modification}).

\subsubsection{\textbf{Step 1: Image blending}}
\label{sec:image-blending}
As discussed earlier, the marked samples should contain the original feature (to preserve visual information), as well as the OOD feature (for provenance purposes). 
Inspired by the common image blending technique~\cite{zhang2018mixup,wenger2022data}, we propose to {blend} the target sample with the OOD feature: 

\begin{equation}
x \oplus (x_{ood}, m)=m \cdot x + (1-m)\cdot x_{ood},
\end{equation}

where $x$ is the original image, $x_{ood}$ is the OOD feature, and $m$ moderates the contribution of different features. 

By using a large $m$, we can largely preserve the high image quality (e.g., the center image in Fig.~\ref{fig:data-modification} is marked with $m=0.7$) while keeping the OOD feature to amplify the model's memorization on the resulting samples. 
We find that merely blending the OOD feature into the target samples ($m=0.7$) can improve the TPR@1\% FPR from 1.28\% to 16.44\%.

\emph{Limitation.} Although the TPR is largely improved, it is still too low. 
We find that this is due to the existence of the original feature, which hinders the model's memorization on the OOD feature. 
We next explain how to mitigate this.

\begin{figure}[t]
  \centering
  \includegraphics[height=.9in]{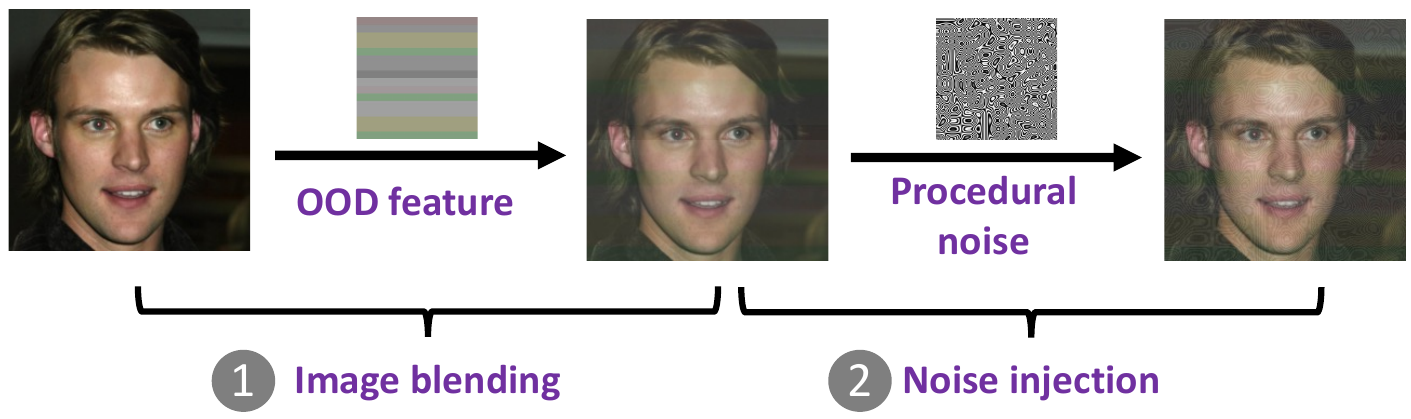}
  \caption{The two-step data marking process: (1) blend the original samples with OOD feature; (2) inject procedural noise. These subtle changes can induce the target samples to be strongly memorized by the model trained on them.}
  \label{fig:data-modification}
  \vspace{-2mm}
\end{figure}

\subsubsection{\textbf{Step 2: Noise injection}}
\label{sec:noise-injection}
Our idea is to inject a small amount of noise to suppress the influence of the original feature, and have the model memorize the OOD feature. 
For this, we draw inspiration from the concept of \emph{adversarial samples}, which
 works by adding imperceptible noise to the inputs and cause them to be misclassified by the model~\cite{liu2016delving,li2020learning,co2019procedural}. 
The perturbation can thus be viewed as the noise that can suppress the influence of the original feature and cause the model to predict the wrong label. 

In our context, we can inject adversarial perturbation into the users' target samples and prompt the model to memorize the OOD feature. 
There are different ways to generate the perturbation. 
Optimization-based approach requires access to some proxy model to generate the perturbation~\cite{liu2016delving,li2020learning}, which can still be challenging for the ordinary users we consider. 

We therefore resort to another optimization-free strategy that does not assume any additional access. 
In particular, Co et al.~\cite{co2019procedural} propose that \emph{procedural noise functions} can be used to generate input-agnostic adversarial perturbation. 
These functions are commonly used in computer graphics to generate different textures and patterns and enrich the image details~\cite{co2019procedural}. 
Co et al. find that the procedural noise patterns share visual resemblance with existing adversarial perturbation patterns, and demonstrate that they can be similarly used to construct adversarial samples. 

This inspires our solution in injecting \emph{perlin noise}~\cite{co2019procedural} to the target samples, and its generation is as follows. 

\begin{equation}
S_{perlin}(x,y) = \sum_{n=1}^{\Omega}p( x \cdot \frac{2^{n-1}}{\lambda_x}, y \cdot \frac{2^{n-1}}{\lambda_y}),
\end{equation}

\begin{equation}
G_{perlin}(x,y) = sin( (S_{perlin}(x,y)) \cdot 2\pi\phi_{sine} ),
\end{equation}

where $\lambda_x, \lambda_y, \Omega, \phi_{sine}$ are to control the noise value at point $(x,y)$. 
$\lambda_x, \lambda_y$ are the wavelengths, $\Omega$ is the number of octaves that contribute to the most visual change, $\phi_{sine}$ is the periodicity of the sine function that creates distinct bands in the image to achieve a high frequency of edges, which can achieve more distinct visual patterns~\cite{co2019procedural}. 
We follow Co et al. to use $L_{\infty}$ norm as the perturbation budget $\delta$ and create random perlin noise~\cite{co2019procedural} for data marking. 

Generating the perlin noise is a lightweight process that does not assume access to a proxy model. 
Despite its simplicity, we find that injecting a small amount of perlin noise (with $\delta=8/255$) is very effective and it can further increase the TPR@1\% FPR from 16.44\% to 57.8\% (a 3.5x improvement). 

Overall, the two-step marking process is able to increase the TPR@1\% FPR from 1.28\% to 57.8\%. 
While the TPR can be further improved by using a smaller $m$ for image blending or larger $\delta$ for noise injection, it would increase the visual distortion (see Appendix~\ref{sec:ablation-diff-marking-param} for an evaluation). 
We thus propose another innovation in the MI process to improve the TPR, and {without} degrading the image quality.

\subsection{Data Provenance via Membership Inference}
\label{sec:mi}
To start with, we note that so far we have been considering MI on a common \emph{instance} basis~\cite{shokri2017membership,carlini2022membership,bertran2024scalable,ye2021enhanced}, which uses the per-instance loss as the signal function value for MI. 
We first explain why this approach would incur low TPR under a low FPR, and then present a specialized MI process to overcome it.

\subsubsection{\textbf{Why instance-based MI suffers from low TPR}}
\label{sec:instance-mi} 
It uses the per-sample loss for MI, which can incur low TPR even if the prediction loss on the target marked samples are very low. 
This is because many non-member samples may also have low loss values. To control the MI process under a low FPR, only a small fraction of member samples with significantly lower prediction loss are recognized as members, thereby resulting in low TPR.

For instance, the left figure in Fig.~\ref{fig:loss-comparison} shows the individual loss values on different  samples, where many non-member samples also yield low prediction loss (the low loss values are shown towards the right side because they are plotted in the logit-scaled form~\cite{carlini2022membership} for  visualization purposes). This leads to a 57.8\% TPR@1\% FPR.

\begin{figure}[t]
  \centering
  \includegraphics[height=1.1in]{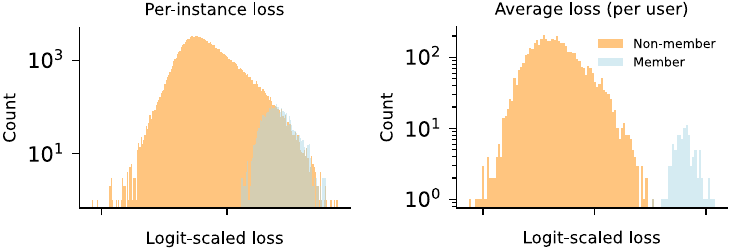}
  \caption{Comparison of using the per-instance loss and the average loss across the samples by each user (each contributes 0.1\% of the training data) as the signal function for MI verification. 
  The former yields 57.8\% TPR@1\% FPR; while the latter has 100\% TPR (our proposal).}
  \label{fig:loss-comparison}
  \vspace{-2mm}
\end{figure} 

\subsubsection{\textbf{Set-based membership inference for reliable data provenance}}
\label{sec:set-based-mi}
To overcome the above, our key insight is that,  although each target user only contributes a small number of samples to the model's training set (e.g., 25 out of 25,000 instances), they can still leverage the \emph{collective} information across their own samples to construct a more reliable signal function for MI, and reduce FPR.

To realize this, 
we start by identifying a pattern that can distinguish the model's behavior on the member and non-member samples, and it is as follows.
For the \emph{non-member samples}, although a few of them may have low loss, the majority of them would incur much higher prediction loss, because these samples are not used for training the model (e.g., see the left of Fig.~\ref{fig:loss-comparison}). 
In contrast, the \emph{member samples} predominantly have very low prediction loss. 

Therefore, we make the observation that \emph{the average loss for the samples by the non-member users are higher than that by the member users}. A visualization for this is on the right of Fig.~\ref{fig:loss-comparison}. 
This leads to our proposal of the \emph{set-based} MI process, which leverages the \emph{average loss} of the small set of target samples by each user as the signal function for MI. We explain the verification process next.

\subsubsection{\textbf{Provenance verification process}}
\label{sec:mi-process}
Algorithm~\ref{alg:mi} outlines the process. 
Let each user possess $k$ samples, where $k$ is a small number compared with the dataset size (e.g., 25 out of 25,000 samples).  
$\mathcal{P}_\text{out}$ denotes a set of non-member data to control the MI process at the low FPR regime (Section~\ref{sec:capability}), which are also marked with some random outlier features and perlin noise (data-marking details in Section~\ref{sec:exp-setup}). 
By comparing the model's behavior on the target data and the data in $\mathcal{P}_\text{out}$, the auditing user $U$ is to detect whether his/her target data are used to train $F$.

The first step is to derive an inference threshold $c$ based on a pre-defined FPR ($\alpha$). 
$\alpha$ is a small number such as 1\% or 0\% to ensure low false positive. 
To derive the threshold $c$, $U$ first computes the empirical loss histogram for data in  the out world $\mathcal{P}_\text{out}$. 
For each non-member user in $\mathcal{P}_\text{out}$, the auditing user ${U}$  first computes the per-instance loss on their data (line 2), and then derives the average loss for each (line 3). 
Line 4  estimates the CDF for the loss distribution, and computes its $\alpha$-percentile to derive a threshold $c$~\cite{ye2021enhanced}.

Next, the auditing user $U$ similarly computes the average loss on his/her samples, and compares it  against the MI threshold $c$. 
If the loss is lower than $c$, the user's data are deemed to be used for training the model (vice versa).

\begin{algorithm}[t] 
\footnotesize 
\begin{flushleft}
\hspace*{\algorithmicindent} \textbf{Input:} $F$: The target DL model;  \\
\hspace*{0.6cm} ${U}$: The auditing user (each user with $k$ target samples); \\
\hspace*{0.6cm} $\mathcal{P}_\text{out}$: Users from the out world (i.e., non-member data);  \\
\hspace*{0.6cm} $\alpha$: False positive rate (FPR) to control the verification process (e.g., 0.1\%); \\
\hspace*{0.6cm} $\mathcal{L}(F, (x,y))$: Loss function (e.g., cross-entropy loss); \\
\end{flushleft}  
  \begin{algorithmic}[1]
    \Function{Data Provenance}{$F, {U}, \mathcal{P}_\text{out}, \alpha, \mathcal{L}$}

    \State $\forall U_i\in\mathcal{P}_\text{out}$, compute $\mathcal{L}(F, U_i), \text{where}~U_i=\{(x^i_j,y^i_j)\}_{j=1}^{k}$ 

    \State $\mathcal{L}_\text{avg} \leftarrow \{\overline{\mathcal{L}(F, U_1)},\cdots, \overline{\mathcal{L}(F, U_n)}\}, n=|\mathcal{P}_\text{out}|  $
    \State $c \leftarrow$ $\alpha$-percentile of the CDF for the loss histogram of $\mathcal{L}_\text{avg}$ 
    \State /* Data provenance based on the inference threshold $c$ */
    \State Compute $\mathcal{L}(F, {U}),  \text{where}~{U}=\{(x_j,y_j)\}_{j=1}^{k}$
    \State $\mathcal{L}^{{U}}_\text{avg}\leftarrow\overline{\mathcal{L}(F, {U})}$

    \If{${\mathcal{L}^{{U}}_\text{avg}} < c$}
    \State ${U}$'s data are used to train $F$ (i.e., member)
    \Else
    \State ${U}$'s data are not used to train $F$ (i.e., non-member)
    \EndIf
    \EndFunction 

  \end{algorithmic}
    \caption{Set-based MI verification process}
    \label{alg:mi}
\end{algorithm}

\section{Evaluation}
\label{sec:eval}

\label{sec:exp-setup}

\textbf{Datasets and model training.}
We consider six common benchmark datasets, including 
CIFAR100~\cite{krizhevsky2009learning}, CIFAR10~\cite{krizhevsky2009learning}, ArtBench~\cite{liao2022artbench}, CelebA~\cite{liu2015deep}, TinyImageNet~\cite{le2015tiny} and ImageNet-1k~\cite{deng2009imagenet}. 
We first consider the first five datasets and the evaluation on the ImageNet-1k dataset is in Section~\ref{sec:imagenet-exp}. 
For CIFAR10 and CIFAR100, we train a WideResNet-28-4 model on 25,000 samples for 100 epochs, with a learning rate of 0.1 (decayed by 5 at epoch 60, 80, 90), a weight decay of 5e-4 and momentum of 0.9. 
For TinyImageNet and ArtBench, we fine-tune an ImageNet-pretrained ResNet-18 model on 25,000 samples. 
For CelebA, we first follow \cite{celeba-code} to create a 307-class subset from the dataset, and then similarly fine-tune an ImageNet-pretrained model on 2,000 samples. 
For fine-tuning, we use a small fine-tuning rate of 0.01 with 30 epochs and a momentum of 0.9. 
In all datasets except CelebA (where we find that fine-tuning on the  entire set yields better accuracy), we use 20\% of the training data as the validation set. 
We also use common data augmentations such as random cropping and horizontal flip.

\begin{figure}[!t]
  \centering
  \includegraphics[height=0.9in]{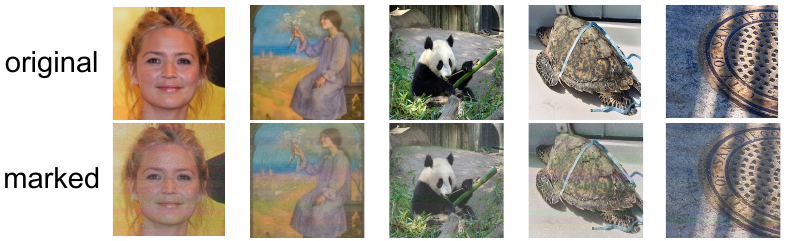}
  \caption{Visualization of the original and marked samples.
  }
  \label{fig:img-exp}
  \vspace{-1mm}
\end{figure}

\textbf{\sysname configuration}. 
We explain the default configurations for the main experiments, and we study different configurations in the ablation study (Appendix~\ref{sec:ablation}). 

As explained earlier, we create outlier features as random color stripes, each with 16 stripes from 11 different common colors. Such a large feature set ($11^{16}$ choices) can support multiple users to generate their own features for data marking, and we use a image blending ratio $m=0.7$. 
Next, we follow Co et al.~\cite{co2019procedural} to generate random perlin noise for each sample with $\delta=8/255$. 
We assume each user possesses a small number of samples from a given class, which constitute 0.1\% of the training set. 

{Fig.~\ref{fig:img-exp} provides additional visualizations of the original and marked samples, and we quantitatively evaluate the visual quality of the images in Appendix~\ref{sec:vis-quality}. 
}

{Unless otherwise stated\footnote{We will use a different metric in Section~\ref{sec:defense} for reasons we will explain there.}, we follow the common practice to use TPR@low FPR for evaluation and we use a 0\% FPR threshold (to minimize false accusations)}. We use 5,000 non-member users for controlling the FPR {and these samples may be collected by sampling random data from the data distribution~\cite{carlini2022membership,bertran2024scalable,ye2021enhanced}}. 
We use all the samples that are not used for training as the non-member set. 
Due to the limited size of the non-member set, the non-member users' samples are drawn randomly from the non-member set (with replacement), and these are similarly marked with random outlier features and perlin noise.

To perform MI, we directly use the cross-entropy loss as the signal function value, which is easy to compute by the users. 
We consider the global-threshold method by Carlini et al.~\cite{carlini2022membership} as the baseline method; and we do not choose the shadow-model-based approaches~\cite{carlini2022membership,bertran2024scalable,ye2021enhanced}. 
The reason is that, even though these methods can achieve better results, they still yield limited increase of MI success (Section~\ref{sec:methodology}); and more importantly, training shadow models is prohibitively challenging for ordinary personal users to carry out (Section~\ref{sec:capability}).  

The closest work to ours in improving the MI success without shadow-model calibration is Tramer et al.~\cite{tramer2022truth}. 
However, their technique assumes the users can manipulate the data labels, which excludes the real-world scenarios where the data labels are assigned by the model creator or an external service~\cite{data-annotation-1,wenger2022data,deng2009imagenet}. 
Though \sysname does not impose such an assumption, for completeness,  we still quantitatively compare it with the method by Tramer et al. (Appendix~\ref{sec:tramer-et-al-comparison}).

\subsection{Results}
\label{sec:res} 
We first evaluate the single-target-user setting, and then the multi-target setting. 

{\textbf{Single-target evaluation}}. 
\label{sec:single-target}
For each experiment, we randomly choose 5 target users from different classes and report the average results. 
Table~\ref{tab:eval-dataset} presents the results. 

\begin{table}[!t]
\centering  
\footnotesize
\caption{Evaluation on different {datasets} (parenthesized numbers show accuracy diff. with the original models). }
\begin{tabular}{ccc}
\hline
Dataset & TPR@0\% FPR & Model accuracy \\
\hline    
CIFAR100 & 100.00 & 66.34 (-0.41)  \\
\hline
CIFAR10 & 100.00 & 90.64 (-0.50) \\
\hline
CelebA & 100.00 & 69.41 (-0.63) \\
\hline
ArtBench & 100.00 & 59.46 (-0.42) \\
\hline
TinyImageNet & 100.00 & 67.22 (-0.40) \\
\hline
\end{tabular}
\label{tab:eval-dataset}
\end{table}

As shown, with \sysname, the users can reliably trace the provenance of their marked data with 100\% TPR@0\% FPR. 
In comparison, without \sysname, performing standard {instance-based} MI on the unmodified target samples cannot reliably identify the member samples under a 0\% FPR threshold and it has 0\% TPR. 
This is because: 1) the unmarked samples cannot be strongly memorized by the model; and 2) the instance-based MI process fails to leverage the collective information across the target samples by each user. 
These two limitations are overcome by \sysname's data marking and set-based MI process; and the detailed ablation study on these two components is presented in Appendix~\ref{sec:ablation}.

Meanwhile, since the target samples only occupy a small fraction of the training set (0.1\%), the models still maintain high performance, and have $<$1\% accuracy difference compared with the ones trained without \sysname.

\begin{table}[t]
\centering  
\footnotesize
\caption{Evaluation on different \emph{models}.}
\begin{tabular}{ccc }
\hline
{Model} & {TPR@0\% FPR } & {Model accuracy} \\
\hline    
WideResNet & 100.00 & 66.34 (-0.41)  \\
\hline
ResNet & 100.00 & 64.49 (-0.69) \\
\hline
ResNext & 100.00 & 63.43 (-0.47) \\
\hline
DenseNet & 100.00 & 67.14 (-0.16) \\
\hline
GoogleNet & 100.00 & 69.09 (-0.33) \\
\hline
SeNet & 100.00 & 65.26 (-0.37) \\
\hline
\end{tabular}
\label{tab:eval-arch}
\end{table}

\begin{table}[t]
\centering  
\footnotesize
\caption{Evaluation on different \emph{training-set sizes}. }
\footnotesize
\begin{tabular}{ccc }
\hline
{Training-set size} & {TPR@0\% FPR } & {Model accuracy} \\
\hline    
25,000 & 100.00 & 66.34 (-0.41) \\
\hline
20,000 & 100.00 & 63.73 (-0.16) \\
\hline
15,000 & 100.00 & 58.99 (-0.83) \\
\hline
10,000 & 100.00 & 50.28 (-0.09) \\
\hline
5,000 & 100.00 & 35.74 (-0.10)  \\
\hline
\end{tabular}
\label{tab:eval-train-size}
\end{table}

\emph{Evaluation on different architectures and training-set sizes.} 
We next consider additional evaluation (using CIFAR100 dataset) on different model architectures (including WideResNet~\cite{zagoruyko2016wide}, ResNet~\cite{he2016deep}, ResNext~\cite{xie2017aggregated}, SeNet~\cite{hu2018squeeze}, GoogleNet~\cite{szegedy2015going} and DenseNet~\cite{huang2017densely}), as well as different training-set sizes (from 5,000 to 25,000). 
We report the results in Table~\ref{tab:eval-arch}  and Table~\ref{tab:eval-train-size}, where we observe a similar trend as before (and thus we omit the discussion).

{\textbf{Multi-target evaluation}}
\label{sec:multi-target} 
We now evaluate how \sysname can support multiple users simultaneously (using CIFAR100 dataset), and  the results are in Table~\ref{tab:multi-target}. 
{We study a large number of target users (more than the number of classes in the dataset), which encompasses the settings where the users are from {different} classes and from the {same} class. }

\begin{table}[htb]
\caption{Evaluation on multiple target users. Supporting more users indicates more specially-marked samples in the dataset, which can lead to higher accuracy drop; but \sysname still maintains high auditing performance.  }
\label{tab:multi-target}
\centering  
\footnotesize
\begin{tabular}{c|cccccc}
\hline
{Num of target users}  &  20 & 50 & 100 & 200 & 300 & 500 \\
\hline
TPR@0\% FPR  & 100.00 & 100.00    & 100.00  & 100.00 & 100.00 & 100.00 \\
Accuracy drop  & -0.44 & -0.56  & -1.39 & -2.32 & -4.47 & -8.93\\ 
\hline
\end{tabular}  
\end{table} 

Supporting multiple users requires the model to memorize more target samples (from different users). 
{However, DL models are known to be capable of memorizing a large corpus of data (e.g., to encode sensitive information~\cite{song2017machine,chen2024method}). 
Likewise, in our case, the model still strongly memorizes the marked samples and enables \sysname to achieve 100\% TPR@0\% FPR even under multiple target users. 
}

Meanwhile, as the number of target users increases, the model also has higher accuracy drop. 
E.g., supporting 100 target users (i.e., 10\% of training set are marked) incurs a 1.39\% accuracy drop, while supporting 500 users has a higher 8.93\% drop. 
This is understandable as supporting more users implies {more} training samples are marked. 
{Supporting more target users will result in a larger accuracy drop - we defer this to future work}}.

\subsection{Robustness to Countermeasures}
\label{sec:defense} 
While Section~\ref{sec:res} demonstrates \sysname's capability in enabling the users to perform reliable data provenance, a deliberate model creator can also employ countermeasures to sabotage \sysname. 

\begin{table}[htb]
\centering 
\footnotesize
\caption{Overview of evaluated countermeasures.
}
\label{tab:defense-summary}
\begin{tabular}{l|l}
\hline
{Defense level} & Defense type \\
\hline
\multirow{2}{5em}{Model level} & \circled{1} MI defense; \circled{2} Adversarial augmentation; \\
                                & \circled{3} Model fine-tuning; \circled{4}  Model pruning \\
\hline
{Input level} & \circled{5} Out-of-distribution and spurious features detection  \\
\hline
{Output level} & \circled{6} Add noise to the output, or return only the label \\ 
\hline
\end{tabular}
\vspace{-1mm}
\end{table} 

To understand this, we evaluate a total of 6 types of countermeasures in the model, input and output level (Table~\ref{tab:defense-summary}), and we use the CIFAR100 dataset. 
For the model-level defenses that require model training, we first consider the single-target setting (similar to Section~\ref{sec:single-target}), and then the multi-target setting. 
For the input- and output-level defenses, we consider the more challenging multi-target setting (100 targets in total).

{
\textbf{Evaluation metric.} 
In this section, we use a different evaluation metric, FPR@high TPR.  
This is because 
using TPR@low FPR (as in our previous evaluation) requires choosing a low FPR threshold and evaluating the corresponding TPR. However, the pre-defined FPR threshold may be too conservative or aggressive. 

We use the results from the DPSGD experiment (Section~\ref{sec:mi-defense}) to explain the issue involved. 
In this example, using a 0.1\% FPR threshold is too conservative as \sysname has the \emph{same} 100\% TPR for the noise multiplier $\sigma=0.01$ and $\sigma=0.03$. Likewise, using a 0\% FPR threshold is too aggressive as \sysname has the \emph{same} 0\% TPR for $\sigma=0.2$ and $\sigma=0.3$. 
In both cases, using a pre-defined TPR may not adequately reflect \sysname's performance changes in different settings. 

To overcome the above issue, we advocate the use of FPR under 100\% TPR. This metric is able to properly compare \sysname's performance under different settings (e.g., Table~\ref{tab:dpsgd}), and hence we use it for the evaluation in this Section, as well as Section~\ref{sec:limitations} and Appendix~\ref{sec:ablation} (for the same reason explained above). 
}

\subsubsection{\textbf{MI defense}}
\label{sec:mi-defense}
We first evaluate several representative privacy defenses for mitigating MI (five in total). 

\textbf{\emph{DP-based defense.}} 
Differentially private (DP) training~\cite{abadi2016deep} is a principled defense that can bound the influence of any training samples to the model, by means of gradient clipping and noise injection to the clipped gradients. 
We adopt the implementation from Aerni et al. to consider the strong DP-SGD baselines that comprise various DP-training techniques such as custom initialization scheme, augmentation multiplicity~\cite{aerni2024evaluations}. We also follow their work to tune hyperparameters for high model utility. 
We use a tight clipping norm of 1, and inject different amounts of noise ($\sigma$). 

The results are in Table~\ref{tab:dpsgd}. 
DPSGD with a loose DP bound is known to be able to reduce the exposure from the \emph{instance-based} MI methods that aim to de-identify some individual training instances~\cite{aerni2024evaluations,carlini2022membership}, and our results also validate this (e.g., in our context, the FPR@100\% TPR under instance-based MI is increased to 47.22\%, with $\sigma=0.01$). 
With that said, \sysname can still overcome the defense via its \emph{set-based} MI process, because the \emph{average loss} of the samples by the member users, though increased due to defense, are still lower than that by the non-member users (average 0.1642 vs. 4.130). 
This avails \sysname to maintain a low 0.01\% FPR@100\% TPR (and the defense already incurs 6.24\% accuracy drop). 
Increasing the amount of noise can further increase the FPR by \sysname, but it also inflicts a larger accuracy drop (undesirable). 

\begin{table}[t]
\centering 
\footnotesize
\caption{Evaluating \sysname against DP-SGD with different amounts of noise injected. 
}
\label{tab:dpsgd}
\begin{tabular}{c|cc}
\hline
{Noise multiplier $\sigma$} & FPR@100\% TPR & Model accuracy \\
\hline
w/o defense & 0.00 & 66.34  \\
\hline
0.01 & 0.01 & 60.10  \\
\hline
0.03 & 0.02 & 57.26  \\
\hline
0.05 & 0.38 & 54.70  \\
\hline
0.2 & 1.07 & 51.56  \\
\hline
0.3 & 3.10 & 50.03  \\
\hline
\end{tabular}
\end{table}

\textbf{\emph{Empirical defense.}} 
Next, we also evaluate two representative defenses that aim to provide strong empirical privacy while preserving model utility: SELENA~\cite{tang2022mitigating} and HAMP~\cite{chen2023overconfidence}. 
We follow the original work to set up both techniques. They both can mitigate the model's memorization (increase the prediction loss) on the marked samples, and with small accuracy drop (2.17\% and 1.42\%). 
However, we find that 
the average loss of the samples by the member users are still much lower than that by the non-member users, and thus \sysname still achieves a low FPR of 0.97\% (on SELENA) and 0.83\% (on HAMP).

Finally, we evaluate two common regularization techniques: \emph{early stopping} and \emph{dropout}. We find that \sysname can still achieve $<0.1\%$FPR@100\% TPR, \emph{unless} under excessive regularization (e.g., aggressively early-stopping the training or using a large dropout rate). However, this would also cause a large ($>10\%$) accuracy drop to the model.

\subsubsection{\textbf{Adversarial augmentation}}
\label{sec:input-pert} 
\begin{figure}[t]
  \centering
  \includegraphics[width=3.35in, height=1.3in]{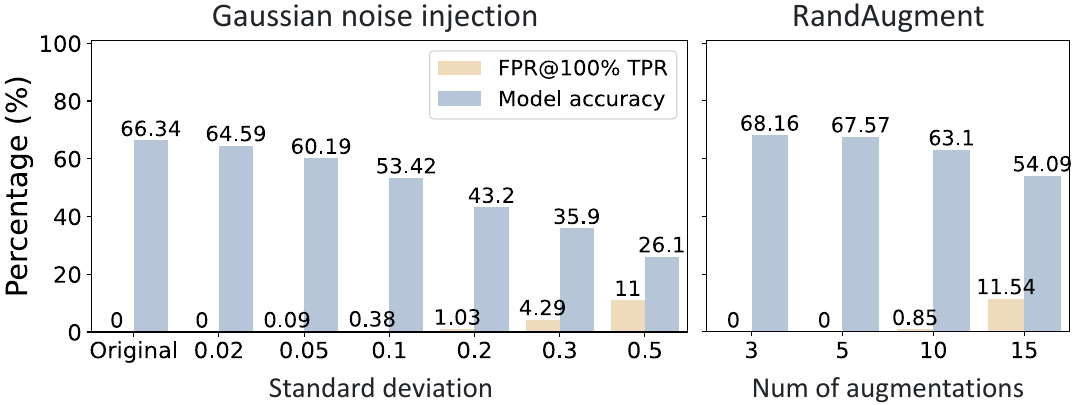}
  \caption{Adversarial augmentation by Gaussian noise injection and RandAugment~\cite{cubuk2020randaugment}. 
  } 
  \label{fig:input-pert} 
  \vspace{-2mm}
\end{figure}

We consider employing additional augmentation techniques during training to mitigate the model's memorization on the marked samples. 
{We study three strategies: (1) injecting Gaussian noise (we inject zero-mean noise with various standard deviation values (0.02$\sim$0.5)); (2) using Mixup~\cite{zhang2018mixup} (we use seven different alpha parameters from 0.1 to 1.5); and (3) using RandAugment~\cite{cubuk2020randaugment} to perform multiple random data augmentations (we consider 3$\sim$15 different augmentations). 

We find that \sysname still maintains 0\% FPR@100\% TPR under Mixup training with different parameters. We report the results for the other two strategies in Fig.~\ref{fig:input-pert}. 
}

When the samples are moderately augmented, in the form of Gaussian noise or multiple random augmentations, \sysname still enables the target users to reliably trace the provenance of their data with low FPR. 

Aggressive augmentation such as applying 15 random transformations to each sample can further increase the FPR to 11.54\%, but it also reduces the accuracy from 66.34\% to 54.09\%, as  such aggressive augmentation severely degrades the image quality and leads to inferior model performance.

\begin{figure}[t]
  \centering
  \includegraphics[ height=1.6in]{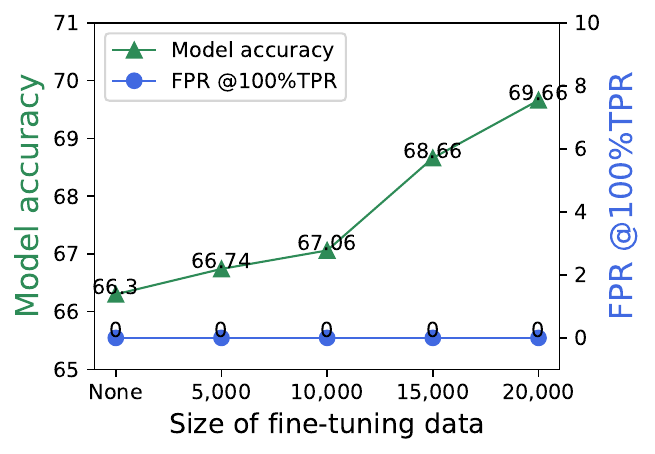}
  \caption{Evaluation with model fine-tuning. Fine-tuning the model with additional data can improve model accuracy (the more data the better), but \sysname consistently maintains high auditing performance regardless. }
  \label{fig:finetune} 
\end{figure}

\subsubsection{\textbf{Model fine-tuning}}
\label{sec:finetune}
{
Fine-tuning is a technique performed on additional data that differs from the training data to improve model performance.
We evaluate how it may be used to mitigate the model's memorization on the target samples. 
We assume an ideal defender with access to an additional set of clean and unmodified samples. We consider multiple settings with different amounts of data (5,000 to 20,000 samples). 
For each setting, we fine-tune the model using various learning rates from 0.1$\sim$0.000001 for 20 epochs, and we report the one with the highest accuracy. 
The results are reported in Fig.~\ref{fig:finetune}.

As shown, using a small set of fine-tuning data (5,000 samples) can increase the model accuracy, and using more data can lead to larger accuracy improvement (understandably so). 
For instance, using 20,000 samples can increase the accuracy to 69.66\%, which translates to a 10\% reduction in the test error. 
Nevertheless, despite its effectiveness in enhancing model accuracy, model fine-tuning still has limited effectiveness in degrading \sysname, which consistently achieves 0\% FPR@100\% TPR in all cases. 
}

\subsubsection{\textbf{Model pruning}} 
\label{sec:pruning}
\begin{figure}[t]
  \centering
  \includegraphics[ height=1.6in]{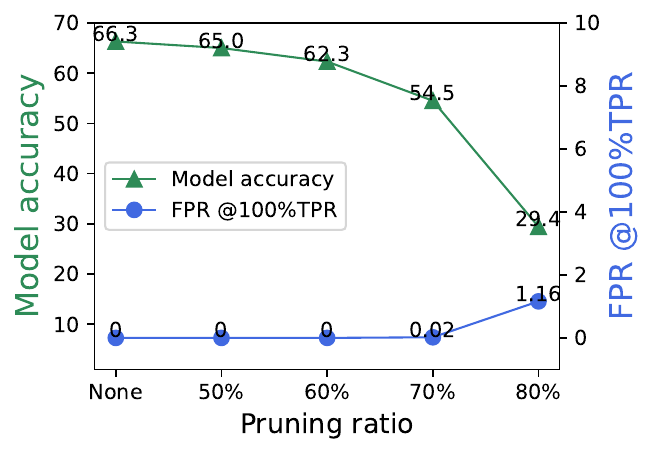}
  \caption{Evaluation with model pruning. \sysname is resilient to moderate model pruning. Aggressive countermeasures can slightly increase the FPR incurred by \sysname, but at the cost of major accuracy degradation.
  }
  \label{fig:pruning} 
  \vspace{-3mm}
\end{figure}

Another related countermeasure we consider is to prune the model's parameters, with the goal of removing the parameters that are associated with the target samples. 
We use the popular pruning strategy by Han et al.~\cite{han2015learning}, which sets the parameters with small absolute values to zero and has minimal impact to the model's performance. 
We evaluate different pruning ratios and the results are shown in Fig.~\ref{fig:pruning}. 

We find that \sysname is highly resilient even when 70\% of the parameters are pruned, where \sysname has only 0.02\% FPR@100\% TPR, but the defense already incurs a 11.8\% accuracy drop. 
More aggressive pruning can slightly increase the FPR, but with a much higher accuracy drop.

\subsubsection{Model-level defenses under multi-target setting}
\label{sec:multi-target-defense}
The previous model-level defenses are evaluated under the single-target setting. We now evaluate a more challenging setting where \sysname needs to support {multiple} users at the same time. We consider 25 target users, and summarize the key findings below.

Among different defenses, we find that \emph{model fine-tuning} stands out as the most potent solution that outperforms other methods in terms of high mitigation performance without accuracy drop. By fine-tuning the model on 20,000 additional  samples, it can increase the FPR@100\% TPR to 42.84\% without accuracy drop (the model still has 67.30\% accuracy).

Hence, unlike the single-target setting where \sysname can maintain high performance, supporting  multiple target users under model-level defenses remains a major challenge by \sysname. We leave the potential improvement of this aspect to future investigation.

\subsubsection{Input detection}
\label{sec:ood} 
Given that \sysname leverages outlier features and perlin noise to the mark the target data, the defender may also perform \emph{out-of-distribution (OOD) detection} or \emph{spurious features detection} on the specially-marked samples. 
We consider both types of countermeasures and we discuss them next.

{\textbf{Unsupervised OOD detection}}. 
We consider eight common detection techniques: Mahalanobis~\cite{lee2018simple}, MSP~\cite{hendrycks2016baseline}, Energy-based OOD~\cite{liu2020energy}, RMD~\cite{ren2021simple}, kNN~\cite{sun2022out}, KLMatching~\cite{hendrycks2022scaling}, VIM~\cite{wang2022vim}, DICE~\cite{sun2022dice}.
We follow the implementation from \cite{kirchheim2022pytorch}, and use a pre-trained ImageNet model to produce the feature representation. 
As before, we assume an ideal defender with access to a small set of clean and unmodified samples (equivalent to 10\% of the training set) for fitting the unsupervised detector, which does not assume any prior knowledge of the potential outlier samples. 

\begin{figure}[!t]
  \centering
  \includegraphics[ height=1.3in]{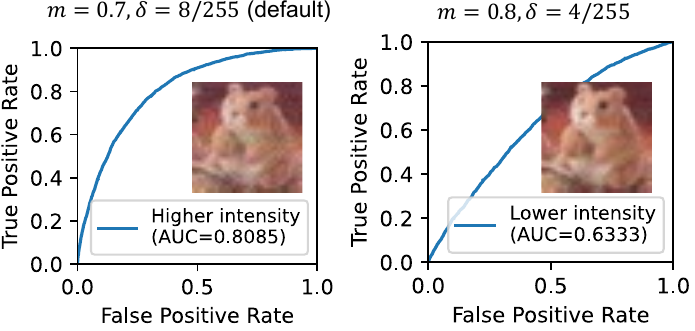}
  \caption{Unsupervised OOD detection. 
  How different data-marking intensities affect the detection performance. 
  \emph{Left}: Samples marked with higher marking intensity are prone to be detected (0.8085 AUC). 
  \emph{Right}: Reducing the marking intensity can effectively reduce the AUC to only 0.6333. 
  }
  \label{fig:ood-unsupervised} 
    \vspace{-1mm}
\end{figure}

We report the results on the left of Fig.~\ref{fig:ood-unsupervised}, where, for brevity, we report only the highest AUC among all detectors. We find that the Mahalanobis-based approach~\cite{lee2018simple} achieves the highest performance with 0.8085 AUC. 
Next, we explain how to (partially) mitigate it.

Our idea is to reduce the amount of distortion applied to the data. 
For this, we conduct an experiment to reduce the intensity of the outlier features (by increasing the image blending ratio $m$ from 0.7 to 0.8) and perlin noise (by reducing the perturbation budget $\delta$ from 8/255 to 4/255). 
The results are shown on the right of Fig.~\ref{fig:ood-unsupervised}, where the highest AUC is reduced from 0.8085 to 0.6333, which translates to a 57\% reduction in the detection performance (over random guessing). 

Naturally, reducing the marking intensity would attenuate the model's memorization on the marked samples, and  increase their prediction loss. 
But we find that the prediction loss on the member samples are still much lower than that on non-members, and thus \sysname can overcome it via the set-based MI process (similar to Section~\ref{sec:mi-defense}), 
 and maintain a 0\% FPR@100\% TPR\footnote{However, there are certain scenarios where the reduced data-marking intensity would lead to lower auditing performance, and an evaluation for this is in Appendix~\ref{sec:ablation-diff-marking-param}.}.

{\textbf{Supervised OOD detection}}. 
Next,  we consider a more knowledgeable defender who is aware of the potential data marking strategy. 
We use the MCHAD method to train a supervised OOD detector~\cite{kirchheim2022multi}, and we consider two scenarios below. 

\begin{figure}[t]
  \includegraphics[height=1.5in]{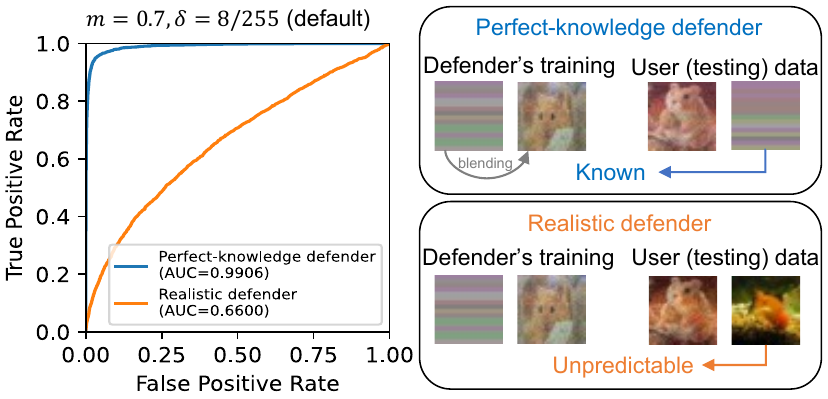}
  \caption{Supervised OOD detection. 
  \emph{Blue} line assumes a defender with \emph{perfect} knowledge of the type of outlier features adopted by the users for data marking.  
  \emph{Orange} line considers a \emph{realistic} defender with knowledge of the possible types of outlier features, but the exact one followed by the users to mark their data is unpredictable. 
  }
  \label{fig:ood-supervised} 
 \vspace{-1mm}
\end{figure}

(1) We first assume a hypothetical defender with \emph{perfect} knowledge of the data marking strategy adopted by the users, i.e., each sample is marked with features consisting of random color stripes and perlin noise. 
The result is in the blue line in Fig.~\ref{fig:ood-supervised}. 
As expected,  the detector  is very capable of detecting the samples that are marked with the same strategy as the one used for training the detector itself (e.g., see the top right illustration in Fig.~\ref{fig:ood-supervised}).  

Though this represents the worst-case scenario against \sysname, it may be difficult to realize in practice, as the outlier features created by the users for data marking can take \emph{diverse} forms (e.g., as features consisting of random color stripes, or features from an OOD dataset - details next), and the exact one used by the users can be hard to predict.

(2) Hence, a more realistic setting is to consider a defender with knowledge of the possible types of outlier features, but is not aware of the exact one adopted by the users to mark their data.  
In this case, the outlier features in the samples used for training the detector are not in the same type as those in users' actual data (e.g., see the bottom right illustration in Fig.~\ref{fig:ood-supervised}). 

Specifically, we train the detector on samples marked with random color stripes, and evaluate it on samples marked with the features from an OOD (TinyImageNet) dataset (this is an effective alternative strategy we study in Appendix~\ref{sec:ablation-diff-ood}).  
The result is in the orange line in Fig.~\ref{fig:ood-supervised}, and the detection performance degrades considerably under such a realistic setting (with only 0.66 AUC).  

Although the defender can attempt to train the detector on samples marked with different types of outlier patterns, with the hope that some of them may overlap with those in users' data, this essentially turns the detection into a cat-and-mouse game, and may not be a desirable practice.

\textbf{Spurious features detection} is another related method. We evaluate a leading technique by  Neuhaus et al.~\cite{neuhaus2023spurious} based on neural PCA components. But we find that it also has limited success in detecting the marked samples (0.6181 AUC).

\subsubsection{\textbf{Output perturbation}} 
\label{sec:output-pert} 
\sysname leverages the model's outputs to perform MI auditing, and we consider two strategies to perturb the model's outputs: (1) injecting Gaussian noise to the outputs; 
and (2) returning only the top-1 prediction label~\cite{choquette2021label,huang2024general}. 
However, we find that \sysname still maintains high auditing success even with the perturbed outputs. 

\begin{figure}[t]
  \centering
  \includegraphics[ height=.9in]{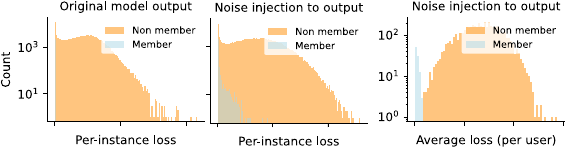}
  \caption{Output noise injection increases the prediction loss of the member samples (from the leftmost to the middle figure), but their loss are still lower than many non-member samples'. 
  Thus, \sysname can leverage the average loss of the samples by each user to achieve a 0\% FPR@ 100\% TPR (the rightmost figure).}
  \label{fig:output-pert} 
\end{figure}

(1) For the first method, we find that injecting Gaussian noise (using zero-mean noise with standard deviation $\sigma$ in 0.5$\sim$5) to the output vectors is not an effective solution, and \sysname can still maintain very low FPR even if the defense has caused severe accuracy degradation. 
E.g., injecting Gaussian noise with $\sigma=3$ degrades the accuracy from 65.32\% to 52.05\%, under which \sysname still achieves a 0\% FPR@100\% TPR. 
We use Fig.~\ref{fig:output-pert} to explain next.

As in the first two figures of Fig.~\ref{fig:output-pert},  injecting noise to the outputs increases the prediction loss on the member samples (and non-members too). 
Hence, more member samples (the blue area in the middle figure of Fig.~\ref{fig:output-pert}) now have similar loss as the non-member samples (undesirable). 

Fortunately, \sysname can still overcome this via its set-based MI process. 
This is because, the prediction loss of the member samples, though increased, are still much lower than that of many non-member samples. 
Thus, there is still a clear difference between the average loss for the member and non-member users' samples (the rightmost figure in Fig.~\ref{fig:output-pert}), from which \sysname is able to achieve a 0\% FPR@100\% TPR.

(2) In the \textbf{label-only} setting, since the class probabilities are not available, we use a simple method to compute a proxy for the prediction loss: we assign a loss value of 0 for the correctly-classified samples and 1 for other samples. 
In this case, all the correctly-classified non-members have the same 0 loss as the member samples. 
However, we find that \sysname can still overcome the defense and achieve a 0.04\% FPR@100\% TPR. 

The reason is that there is a major accuracy difference between the member and non-member users' samples: 100\% vs. 22.02\% (the latter yield low accuracy as they are marked with the random outlier features and perlin noise, but they are \emph{not} present in the training set). 
Thus, the ``average loss'' for the samples by the member users are still lower than that by the non-member users. 
This enables \sysname to succeed even under the label-only setting.

\subsection{Evaluation on ImageNet Training}
\label{sec:imagenet-exp}
This section evaluates \sysname under the full-sized ImageNet-1k training (over 1.28 million training samples). 
We train a ResNet50~\cite{he2016deep} and Swin-Transformer~\cite{liu2021Swin}, and they have 75.09\% and 80.29\% accuracy respectively (both with <1\% accuracy drop). 
We consider the challenging setting of supporting multiple users in different classes (1,000 target users in total). Each user contributes 125 samples to the training set, which is $<$0.01\% in proportion. 

We find that even under the large-scale training setting, the marked samples can still be strongly memorized by the models, which enables \sysname to achieve 100\% TPR@0\% FPR.

\begin{figure}[t]
  \centering
  \includegraphics[ height=1.5in]{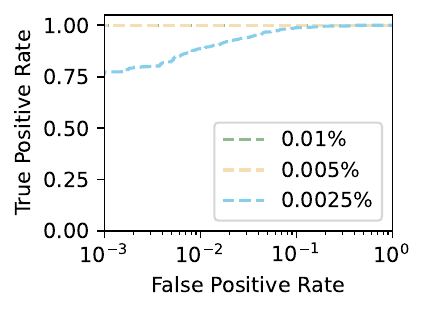}
  \caption{Evaluating \sysname under full-sized ImageNet training with different portions of samples for data marking. 
  }
  \label{fig:imagenet} 
  \vspace{-2mm}
\end{figure}

Next, we evaluate how the performance varies with \emph{lesser} samples for marking (using ResNet50): we consider 63 and 32 samples by each user, which amount to $\sim0.005\%$ and $\sim0.0025\%$ of the training set. 
The results are in Fig.~\ref{fig:imagenet}.

\sysname can still maintain 100\% TPR@0\% FPR, when the marking ratio is reduced from 0.01\% to 0.005\%. 
However, when the marking ratio is further reduced to 0.0025\%, \sysname experiences lower auditing performance - 
this trend is similar to that in our ablation study in Appendix~\ref{sec:ablation-two-step-marking} and thus we defer the detailed discussion to Appendix~\ref{sec:ablation-two-step-marking} (due to space constraints). 
With that said, even under such a challenging case, \sysname still achieves reasonably good performance with $>$75\% TPR@0.1\% FPR (the blue line of Fig.~\ref{fig:imagenet}). 
\emph{To the best of our knowledge, \sysname is the first technique that can scale to support reliable data auditing in large-scale training setting.}

\section{Discussion}
\label{sec:limitations} 

{In this section, we first discuss how our assumption on the access to a set of non-member samples can be relaxed, and then discuss the limitations of \sysname. 
}

\subsection{(Relaxed) Assumption on Non-member Set} 

\textbf{Do non-member samples need to be definitely outside the training set?}  
Recall that \sysname follows a \emph{common} assumption~\cite{carlini2022membership,bertran2024scalable,wenger2022data,ye2021enhanced} that users have access to a set of non-member samples (for calibrating the MI process under low FPR regime). 
However, in practice, it can be challenging for users to determine whether these non-member samples are {definitely not} used for training the target model. 
For instance, an auditing user may collect some random facial images from the Internet as non-member samples, but it is \emph{possible} that these samples already have been used for training the target model. 

To understand how this may affect \sysname's performance, we study a hypothetical scenario where all the non-member samples (used for data auditing) have been used for training the target model. 
Even under such a challenging case, we find that \sysname can still maintain reliable auditing performance with 100\% TPR@0\% FPR. We explain the reason below.

Recall that \sysname performs data marking on non-members using random outlier features. These random features are \emph{highly unlikely} to be used for training (e.g., among $10^{16}$ possible random color stripe features), and thus they  significantly increase the prediction loss of non-member samples, e.g., reducing their accuracy from $>$99\% (this is high because these non-member samples are used for training) to $<$30\% after data marking. 
Consequently, the marked non-member samples still exhibit much higher prediction loss than marked member samples, from which \sysname still maintains 100\% TPR@0\% FPR for reliable data auditing. 

Overall, we find that with \sysname, users only need to collect some data from the underlying population, and these data can be \emph{any} data; they do not necessarily have to be outside the training set of the target model. This demonstrates the broader practicality of \sysname.

\subsection{Limitations}
\label{sec:limitations}
Next, we discuss the three main limitations by \sysname. 

{\textbf{\circled{1}}} 
\sysname works by marking a small number of target samples to be strongly memorized by the model, and detecting whether the model memorizes these samples. 
However, a malicious developer may seek to jeopardize \sysname's performance, by using only a \emph{subset} of data from each user for training, so that not all target samples are memorized by the model.

\begin{table}[t]
\centering 
\footnotesize
\caption{Evaluation on the partial inclusion of user data in model training. 
With lesser data (50 samples per user) included for training, 
the average loss on the entire set of samples by each target user increases. This in turn (1) increases the FPR incurred by \sysname, and (2) leads to larger accuracy drop due to lesser data for training.
}
\label{tab:partial-training}
\begin{tabular}{c|c|c}
\hline
\% of user samples &  \multirow{2}{7em}{FPR@100\% TPR} &   \multirow{2}{7em}{Accuracy drop} \\
used for model training   &   &   \\
  \hline
100\% (50/50) & 0.00   & 0.00  \\
90\% (45/50) & 0.00  & 0.12  \\
70\% (35/50) & 0.00  & 4.83  \\
50\% (25/50) & 0.34  & 12.34 \\
40\% (20/50) & 0.84  & 14.76 \\
30\% (15/50) & 2.42  & 21.37 \\
\hline
\end{tabular}
\end{table}

To understand how this affects \sysname, we perform an experiment where we assume the model owner collects 50 samples from each user (0.2\% of the training set), and he/she includes only a subset of samples from each user for training. 
Table~\ref{tab:partial-training} shows the results, where we consider five target users (we observe a similar trend when considering the support for more target users). {We use FPR@100\% TPR as the metric, as explained earlier in Section~\ref{sec:defense}.}  

The FPR incurred by \sysname grows as the number of marked samples used for training decreases. 
This is because those samples that are not used for training would yield higher prediction loss, which in turn increases the average loss on the entire set of marked samples, and affects \sysname. 
On the other hand, while using a subset of user data can degrade \sysname's performance, it also results in lesser data for model training and causes an accuracy drop (third column in Table~\ref{tab:partial-training}). 

Overall, we find that \sysname can still maintain $<0.5\%$ FPR@100\% TPR, even when only 50\% of the user data are used for training. In this case, the model already suffers from a 12\% accuracy drop. 
Using lesser amounts of user data for training can further degrade \sysname's auditing performance, but it will also result in a larger accuracy drop (undesirable).

\textbf{\circled{2}} 
Despite our best effort in keeping the targeted changes from severely  distorting the marked data, the artifacts created by \sysname are still visible under close inspection (e.g., Fig.~\ref{fig:img-exp}). 
However, performing manual inspection to filter out the specially-marked samples by \sysname can still be a challenge, particularly under the large-scale training setting (e.g., with millions of images). 
With that said, creating more subtle targeted changes 
 still remains an important venue for future investigation.

\textbf{\circled{3}} {Finally, since \sysname targets the model's propensity in memorizing data, in principle, {any} effective measures that can mitigate the model's memorization on training data would be able to degrade 
the performance by \sysname. 
Indeed, our evaluation in Section~\ref{sec:defense} {comprehensively considers multiple measures applied to model, input, and output levels};  we find that though some  approaches such as differential privacy can be configured to degrade the auditing success achieved by \sysname, they also result in utility loss to the model (undesirable). 

Given that building high-performance models often represents the model creators' foremost interest, the development of tools like \sysname can prompt the practitioners to reconsider the choice of resorting to other legal data acquisition means and building high-performance models using the authorized data. 

}

\section{Conclusion and Future Work}
\label{sec:conclusion}
This work presents \sysname, a practical data provenance tool that can support ordinary users to audit whether their specific data are used to train deep learning models without their permission. \sysname consists of a lightweight data marking process that can mark the target data with small and targeted changes, and a high-power set-based membership inference process that can reliably trace the provenance of the target data. 
By merely marking a small fraction of samples (0.005\%$\sim$0.1\% of the dataset), \sysname effectively empowers the users to reliably audit the usage of their data (average 100\% TPR@0\% FPR), and it is also scalable to the large-scale (ImageNet) training setting. 
If their data are found to be misused, the users can take legal action or request the ``right to be forgotten'' in accordance with privacy regulations such as GDPR~\cite{gdpr} (e.g., based on techniques on  machine unlearning~\cite{bourtoule2021machine}).
  
Finally, we discuss future work directions. 
First, while \sysname retrofits the conventional membership inference (attack) to enable responsible AI, it also faces the risk of being misused by the malicious party. 
Notably, the data marked with \sysname would now become easier to be de-identified by \emph{any} party with access to them.  
However, \sysname should be applied only when the users do \emph{not} want their data to be used for training DL models, in which membership privacy may not be of the users' concern. 
By contrast, if the users are willingly sharing their data for model training, 
they should refrain from using \sysname to avoid unnecessary membership exposure. 
Future studies can explore potential usage cases where \sysname is applicable \emph{and} protecting the membership privacy of the marked samples is necessary, and consider the further adaptation of \sysname to cater to such a need. 

Second, future work can explore the potential of extending \sysname to other domains like generative models. 
Large generative models are often trained on massive amount of data collected from the Internet~\cite{rombach2022high,ramesh2022hierarchical}, and are the frequent targets where unauthorized use of personal/copyrighted contents arises~\cite{ai-art-lawsuit,midjourney}. 
Thus, 
future work can investigate whether the large generative models  would also be prone to memorize the samples marked with \sysname under the billion-scale training environment.


\bibliographystyle{ACM-Reference-Format}
\bibliography{\jobname}

\appendix

\section{Quantitative Evaluation of Image Visual Quality}
\label{sec:vis-quality}
{A key goal of our work is to preserve high visual quality. 
While we provided some visualization examples earlier in Fig.~\ref{fig:img-exp}, in this section, we conduct experiments to quantitatively assess the visual quality of the images marked by \sysname. 

We use three commonly used metrics: (1) Mean Squared Error (MSE), (2) Structural Similarity Index Measure (SSIM) and (3) Learned Perceptual Image Patch Similarity (LPIPS) to assess visual quality. 
MSE measures the average squared difference between corresponding pixel values of two images. SSIM is commonly used to assess the similarity between two images, considering changes in structural information, luminance, and contrast~\cite{wang2004image}. LPIPS computes the distance between feature representations of two images extracted from a pre-trained neural network~\cite{zhang2018unreasonable}. This approach captures more nuanced differences in texture and structure, aligning more closely with human perception of image quality. 

We use the built-in SSIM implementation from TorchMetric~\cite{torchmetrics}. We adopt the LPIPS implementation from the original authors~\cite{zhang2018unreasonable}, and we follow their technique to use an  AlexNet for extracting the feature representations. We use 5,000 random images for evaluation under three settings: \circled{1} images marked with full \sysname (with OOD feature and procedural noise); \circled{2} images blended with OOD feature (without procedural noise); and \circled{3} the strawman approach (which replaces the target images with OOD feature). 
We report the results in Table~\ref{tab:vis-quality}. 

\begin{table}[t]
\centering  
\footnotesize
\caption{Quantitative evaluation of image visual quality. \circled{1} Images marked with full \sysname (with OOD feature and procedural noise); \circled{2} images blended with OOD feature (without procedural noise); and \circled{3} the strawman approach (which replaces the target images with OOD feature).}
\begin{tabular}{c|ccc }
\hline
Settings & MSE & SSIM & LPIPS \\
\hline    
\circled{1} & 0.0077 & 0.8424 & 0.0244  \\
\hline
\circled{2} & 0.0072 & 0.8900 & 0.0173 \\
\hline
\circled{3} & 0.0814 & 0.0923 & 0.4076 \\
\hline
\end{tabular}
\label{tab:vis-quality}
\end{table}

For full \sysname (\circled{1}), the average SSIM is high (0.8424), and the LPIPS and MSE are low (0.0244 and 0.0077), indicating high similarity between the marked and original images.
Removing the procedural noise (\circled{2}) can improve the visual quality, under which the MSE is reduced to 0.0072, the SSIM is increased to 0.8900 and LPIPS reduced to 0.0173. 
Compared with \circled{1} and \circled{2}, the strawman approach (\circled{3}) has the highest MSE and LPIPS and the lowest SSIM, which is understandable as it replaces the target image with OOD features. 
Overall, our results quantitatively validate that the images marked by full \sysname can still preserve high visual quality under different metrics. 
}

\section{Ablation Study}
\label{sec:ablation}

\sysname consists of a data marking and set-based MI process, and we conduct a detailed ablation study into these two in Appendix~\ref{sec:ablation-data-modification} and \ref{sec:ablation-mi}, respectively. 
{As explained earlier in Section~\ref{sec:defense}, we use FPR@100\% TPR as the evaluation metric.} 

\subsection{Ablation of the Two-step Data Marking}
\label{sec:ablation-data-modification}
We first evaluate the effectiveness of the two data-marking components (Appendix~\ref{sec:ablation-two-step-marking}). Meanwhile, we also study how \sysname's performance varies under different number of target samples contributed by the users. 

Next, we evaluate how different data-marking intensities may affect \sysname's performance (Appendix~\ref{sec:ablation-diff-marking-param}). 
Finally, we explore alternative data marking methods in Appendix~\ref{sec:ablation-diff-ood}.

\subsubsection{Investigating the two-step data marking process}
\label{sec:ablation-two-step-marking} 
This section compares the performance of different variants of \sysname, including 
(1) the full \sysname with image blending and noise injection; 
variant (2) with image blending only; 
variant (3) with noise injection only; and 
variant (4) without any data marking. 

We consider a total of 100 target users, and each contributes different number of samples (5 to 15) to the training set (25,000 instances). Table~\ref{tab:ablation} shows the results, and there are two main findings. 

\begin{table}[t]
\caption{Ablation study of \sysname's two-step data marking process: \circled{a}: Image blending with OOD feature; \circled{b}: Perlin noise injection. Each \emph{column} reports the performance under a specific number of target samples contributed by the users (5$\sim$15 samples by each). 
  }
\label{tab:ablation}
\centering  
\footnotesize
\begin{tabular}{|c|ccc|}
\hline
\multirow{3}{4em}{Approach}  & \multicolumn{3}{c|}{FPR@100\% TPR} \\
        \cline{2-4}  & 15 samples & 10 samples  & 5 samples  \\
          & (0.06\% dataset) & (0.04\% dataset) & (0.02\% dataset) \\
\hline
\circled{a} + \circled{b} & 0.00\% & 0.18\% & 0.64\% \\
\hline
\circled{a} & 0.10\% & 0.96\% & 4.62\% \\
\hline
\circled{b} & 0.06\% & 0.40\% & 1.8\% \\
\hline
None  & 0.34\% & 1.54\% & 7.18\% \\
\hline
\end{tabular}  
\end{table} 

(1) \textbf{The two-step data marking process is vital for \sysname's high auditing success}. 
The full technique with the two-step process consistently achieves lower FPR than the two other variants with only a single-step marking (second and third row in Table~\ref{tab:ablation}). 
The last variant (the bottom row in Table~\ref{tab:ablation}) considers the set-based MI process on the original target samples without any data marking, which suffers from even higher FPR. 
E.g., under 5 target samples per user, the full technique has only 0.64\% FPR vs. 7.18\% by the one without data marking. 

Meanwhile, the performance difference between different approaches in Table~\ref{tab:ablation} shrinks as the number of target samples contributed by the users increases. 
Thus, there is a low FPR even when  auditing the original target samples (without data marking), e.g., 0.34\% FPR@100\% TPR on the setting with 15 target samples per user. 
This is in fact due to the effectiveness of the proposed set-based MI process\footnote{The ablation study on \sysname with and without the set-based MI process is in Appendix~\ref{sec:ablation-mi}.}
{However, this is \emph{not} enough and using the two-step data marking still achieves better performance. The performance advantage of performing data marking is even more profound under the more challenging cases with limited user samples (third and fourth column in Table~\ref{tab:ablation}). 
}

(2) \textbf{Performance variation under different number of target samples by the users}. 
Table~\ref{tab:ablation} also shows \sysname's performance degrades as the number of target samples by the users reduces (this trend is also similar to our earlier evaluation of the ImageNet training experiment in Section~\ref{sec:imagenet-exp}). There are two reasons. 

First, recall that in \sysname, each user applies a random outlier feature to mark their data, and with lesser data for marking, the outlier feature appears in fewer samples. 
Thus, the model's memorization on the marked samples (that contain the outlier feature) subsides. This can be reflected from the higher prediction loss on the samples, and we validated this. 

Secondly, the limited size of target samples also reduces the amount of information available to the set-based MI process, which can result in a lower auditing performance as well. 
With that said, even in the challenging case where each user can only mark 5 samples (0.02\% of the dataset), the full \sysname still incurs a reasonably low FPR of 0.64\% with 100\% TPR.

\begin{figure}[t]
  \centering
  \includegraphics[width=3.4in, height=1.1in]{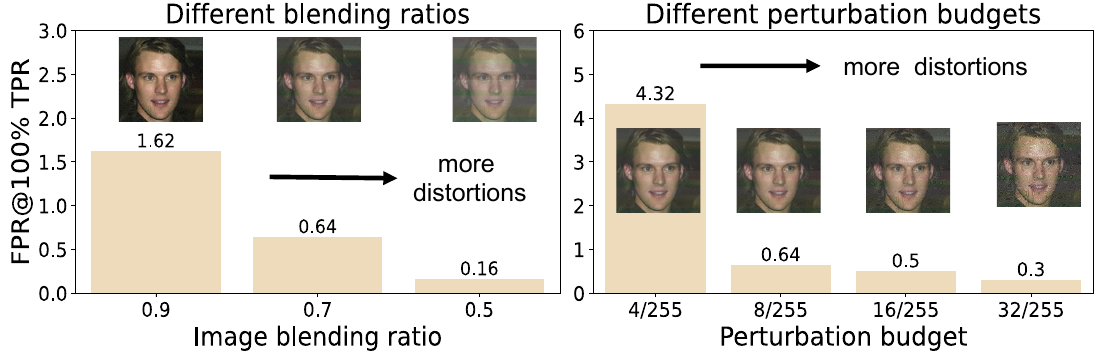}
  \caption{Performance evaluation under different image blending ratios (\emph{left}) and noise perturbation budgets (\emph{right}) (5 target samples per user). 
  }
  \label{fig:diff-param} 
\end{figure}
\subsubsection{Data marking with different parameters}
\label{sec:ablation-diff-marking-param}
There are two parameters: the blending ratio $m$ and noise perturbation budget $\sigma$. 
Our main experiments use $m=0.7$ and $\sigma=8/255$, and we study how different parameters may affect \sysname's performance. 
The results are in Fig.~\ref{fig:diff-param}, where we consider a more challenging case with 5 target samples per user, because the performance difference between different marking intensities is less distinctive when there are more target samples by each user (e.g., 15). 

In both figures of Fig.~\ref{fig:diff-param}, higher marking intensity indeed enhances the ability of the marked samples to be memorized by the model, and contributes to the better performance by \sysname. 
This illustrates that the users can moderate the balance between the amount of distortion applied to their data and the degree of auditing performance obtained from the resulting data.

\begin{figure}[t]
  \centering
  \includegraphics[ width=3.1in, height=1.3in]{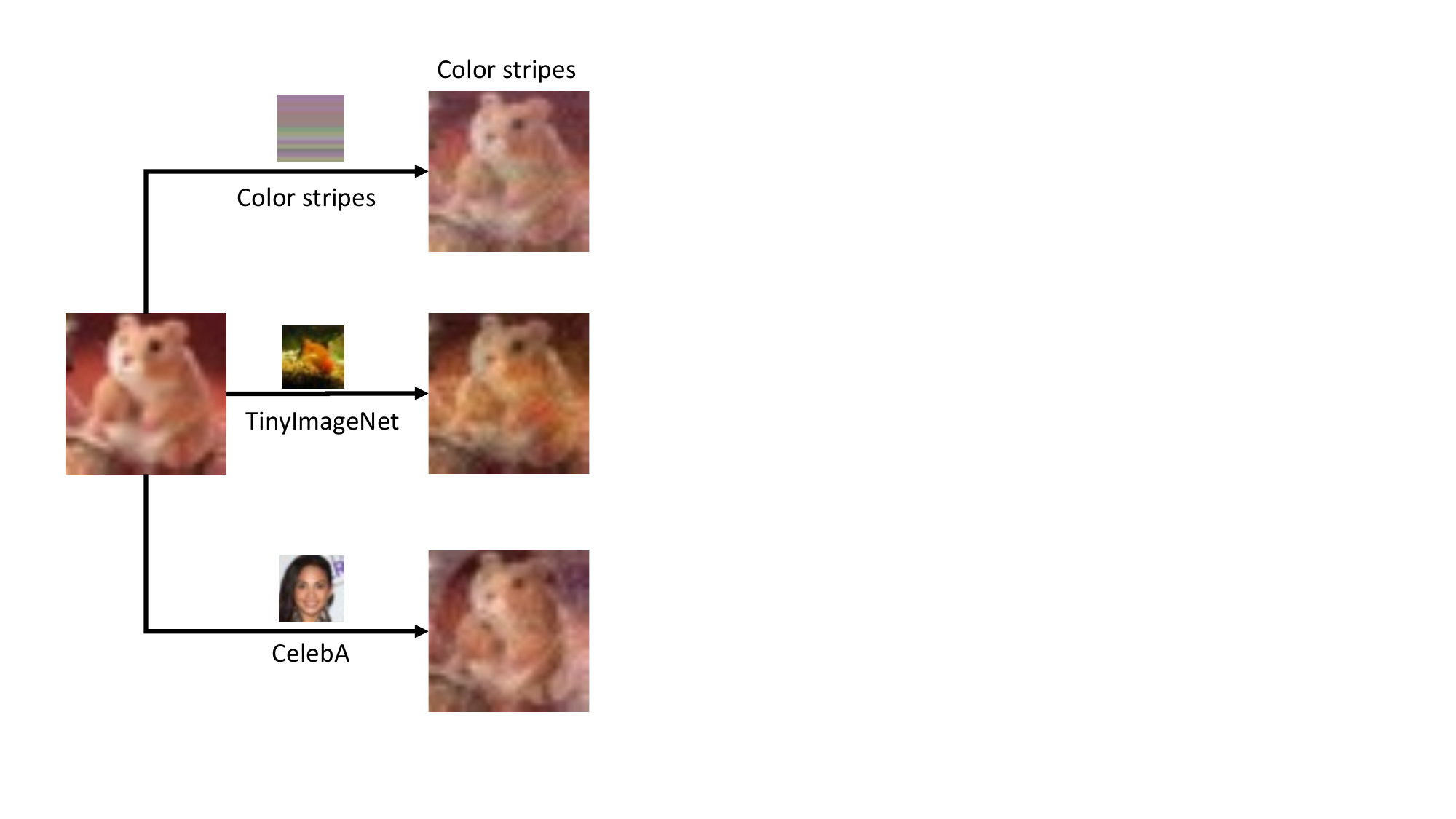}
  \caption{\sysname maintains similarly high performance when using different types of outlier features for data marking, such as samples with random color stripes, samples from an OOD dataset (TinyImageNet and CelebA).  }
  \label{fig:diff-ood-res} 
\end{figure}

\subsubsection{Alternative data marking methods} 
\label{sec:ablation-diff-ood} 

\emph{(1) Alternative outlier feature generation.}
In addition to constructing outlier features as samples consisting of random color stripes, we consider alternative strategies by using samples from an OOD dataset. 
Specifically, we use random samples from the TinyImageNet and CelebA dataset, which are blended into the CIFAR100 samples. 
The results are shown on the left of Fig.~\ref{fig:diff-ood-res}, and a visualization on the right. 

As shown, the resulting samples can be similarly memorized by the model and enable \sysname to maintain high provenance success. 
E.g., in the last two rows on the left of Fig.~\ref{fig:diff-ood-res} (where each user marks 20 or 25 samples), using different types of outlier features can similarly avail the users to achieve 0\% FPR@100\% TPR. 
This offers the flexibility for the users to choose different types of outlier features to mark and protect their data.

\emph{(2) Alternative noise injection.} We compare injecting perlin noise vs. uniform random noise. We find perlin noise yields better results (details omitted), because perlin noise is a stronger adversarial noise~\cite{co2019procedural}, and thus more capable of inducing the model to memorize the marked samples. 
Other alternatives such as using other noises like gabor noise~\cite{co2019procedural} may also be explored in future studies.

\subsection{Ablation of the Set-based MI Verification}
\label{sec:ablation-mi}
This section compares the set-based MI with the common instance-based MI process.  
Table~\ref{tab:ablation-MI} shows the results where the target samples are marked with \sysname (and we observe a similar trend when comparing both approaches on the original samples {without} data marking). 
As in Table~\ref{tab:ablation-MI}, the proposed set-based MI process consistently outperforms the instance-based MI method. 
E.g., when each target user can  mark only 15 samples (the second column in Table~\ref{tab:ablation-MI}), they can achieve 0\% FPR@100\% TPR via the set-based MI process, while the instance-based MI incurs 20.47\% FPR. 

\begin{table}[!t]
\caption{Comparing the instance-based and the proposed set-based MI process. The latter consistently outperforms the former method by leveraging the average loss across the target samples by each user to perform MI auditing. 
}
\label{tab:ablation-MI}
\centering  
\footnotesize
\begin{tabular}{|c|ccc|}
\hline
\multirow{3}{5em}{Approach}  & \multicolumn{3}{c|}{FPR@100\% TPR} \\
      \cline{2-4}    & 10 samples  & 15 samples  & 25 samples\\
          & (0.04\% data) & (0.06\% data) & (0.10\% data) \\
\hline
Instance-based MI & 36.09 & 20.47 & 7.97 \\
Set-based MI  & 0.18 & 0.00 & 0.00 \\
\hline
\end{tabular}  
\end{table} 

Meanwhile, the FPR by the instance-based MI approach reduces as the number of marked samples contributed by the users increases (this trend is similar to that in Table~\ref{tab:ablation} discussed earlier). 
Nevertheless, even if each user can mark 25 samples, the instance-based MI method still incurs 7.97\% FPR@100\% TPR, while the set-based MI has 0\% FPR.

\section{{Comparison with Wenger et al.~\cite{wenger2022data}}} 
\label{sec:wenger-comp} 
Wenger et al. propose a technique to audit the unauthorized use of data in training DL models. It first injects a target spurious feature into the user data, and then detects whether the model has associated the injected feature with a target class label. 
To detect, the users can separately overlay the target feature and some non-target features to an auxiliary set of samples (outside the target class), and then check whether the target feature causes a  slightly higher probability shift to the target class in those samples, compared with the non-target features. 
For instance, assume the target class is ``cat''. The user may compare the cat class probability on some ``dog'' images that are overlaid with the target and non-target features (e.g., 10\% vs. 1\%), and detect potential data misuse. 

In the following, we first discuss a notable issue we found in their work, which leads to a significant underestimate of false positives by their technique; and then further compare both techniques.

\textbf{Analyzing the problematic evaluation in Wenger et al~\cite{wenger2022data}. } 
In their work, a testing feature should be detected as being used for training, if it causes higher probability shift than the \emph{non-target} feature - this procedure should be the \textbf{same} regardless of whether the testing feature has actually been used for training or not (i.e., true or false positive). 
Unfortunately, this is not the case in \cite{wenger2022data}. 

\begin{algorithm}[t] 
\footnotesize 
\begin{flushleft}
\hspace*{\algorithmicindent} \textbf{Input:} $T_{in}$: the \emph{true target} spurious feature used for model training;  \\
\hspace*{0.6cm} $T_{out}, T_{out}'$: two random \emph{non-target} features (\emph{not} used for training); \\
\hspace*{0.6cm} /* Assume the prob shift by $T_{in}, T_{out}, T_{out}'$ to be 10\%, 5\% and 1\% */
\end{flushleft}  
  \begin{algorithmic}[1]

    \item[\hspace*{0.5cm} /* \cmark~Compare $T_{in}$ with $\textcolor{teal}{T_{out}'}$ to check (true) positive */]
    \IIf{Prob shift by $T_{in} > \textcolor{black}{T_{out}}'$ } true positive += 1 \EndIIf

    \item[]
    \item[\hspace*{0.5cm} /* \xmark~\cite{wenger2022data} invert the order of Line 1 to check (false) positive */ ]
    \IIf{Prob shift by  $T_{out}' > \textcolor{purple}{T_{in}}$ } false positive += 1 \EndIIf  \hspace*{0.2cm} 
    \item[\hspace*{0.5cm} /* This would report \emph{no} false positive */ ]

    \item[]
    \item[\hspace*{0.5cm} /* \cmark~Sample a new $T_{out}$ and compare it with $\textcolor{teal}{T_{out}'}$ for FP */]
    \IIf{Prob  shift by $T_{out} > \textcolor{black}{T_{out}'}$ } false positive += 1 \EndIIf \hspace*{0.2cm} 
    \item[\hspace*{0.5cm} /* This would report a false positive */ ]
    \item[\hspace*{0.5cm} /* \xmark~But it still \textcolor{purple}{cannot} control the test under a specific low FPR */ ]

    \item[]
    \item[\hspace*{0.5cm} /* \cmark~A \textcolor{teal}{principled} way to satisfy the above requirement */ ]
    \State Estimate the prob shift by sampling a large number of non-target features
    \item[\hspace*{0.5cm} /* $\alpha$ can be a small number such as 0.01\% (for the low FPR) */ ]
    \State $c \leftarrow$ $(1-\alpha)$-percentile of the CDF for the above estimated prob shift 
    \If{Prob shift by $T_{test} > \textcolor{teal}{c}$} 
    \State \textbf{if} {$T_{test}$ belongs to $T_{in}$} \textbf{then} true positive += 1 
    \State \textbf{else} false positive += 1 
    \State \textbf{end if}
    \EndIf

  \end{algorithmic}
    \caption{Illustrating the problematic evaluation in measuring \emph{false positive} by \cite{wenger2022data}, and how to fix it. }
    \label{alg:wenger}

\end{algorithm}

We use Alg.~\ref{alg:wenger} as an example to explain in the following. 
Assume $T_{in}$ is the true target feature that causes a probability shift of 10\%, and $T_{out}$ 5\% and $T_{out}'$ 1\% ($T_{out}$ and $T_{out}'$ are two random non-target features that are  \emph{not} used for training). 
Line 1 in Alg.~\ref{alg:wenger} compares $T_{in}$  with $T_{out}'$ to check true positive. 

To compute false positives (Line 2), however, Wenger et al. \cite{wenger2022data} simply \emph{invert} the order of Line 1, which essentially compares the probability shift between the non-target feature $T_{out}'$ and the \emph{true target} feature ($T_{in}$). 
This would report \textit{no} false positives (because $T_{out}'$ has a lower probability shift than $T_{in}$), but it is not the correct way to compute false positive.

Instead, a fair evaluation should: (1) sample a \emph{new non-target} feature ($T_{out}$); and (2) compare its probability shift relative to \emph{another non-target} feature ($T_{out}'$)\footnote{In fact, the order of $T_{out}$ and $T_{out}'$ in Line 3 of Alg.~\ref{alg:wenger} can be interchangeable, as they are \emph{both} random features that are not used for training. We use the given order only to illustrate how Wenger et al. \cite{wenger2022data} could overlook a false positive in their original detection.}. 
This is illustrated in  Line 3 of Alg.~\ref{alg:wenger} (which is basically the same as Line 1 for detecting true positive). 
In such a rectified procedure, the technique  \textit{would have} a false positive, because $T_{out}$ has a higher probability shift than $T_{out}'$.

\emph{Validation.} To evaluate how this affects their results, we use the original implementation from~\cite{wenger2022data}.  
We use CIFAR100 dataset with 50 target classes and 50 non-target features for comparison; and otherwise follow the the same parameters they used in their experiment~\cite{wenger2022data}. 
We first validate that we can obtain similar results to their reported performance, with 99.63\% TPR and 0\% FPR\footnote{We identified a bug in their code (an issue related to  pass-by-reference in Python that results in the code using a different set of auxiliary data for marking the target and non-target features), and we fixed it in our evaluation. We have also notified the authors about this bug in their official code repository using an anonymous account.}. 

To re-evaluate the FPR following Line 3 in Alg.~\ref{alg:wenger}, we compare the probability shift between two disjoint sets of non-target features ($T_{out}, T_{out}'$). Because both feature sets are not used for training, the FPR should be similar if we compare $T_{out}$ Vs. $T_{out}'$ or $T_{out}'$ Vs. $T_{out}$ (i.e., their order is interchangeable). 
We indeed did observe this (with 37\% and 34.6\% FPR). On average, their technique incurs a 35.8\% FPR. 

Wenger et al. \cite{wenger2022data} do mention that comparing two sets of non-target features is  equivalent to random guessing, and we also observe the same. 
However, this ``indistinguishablity'' applies only when attempting to distinguish two \emph{set} of features; if we consider two random \emph{individual} features, it is  likely that one feature will cause a higher/lower probability shift than the other. 
This is the exact issue with Wenger et al.'s method, which performs each test on a  \emph{feature-to-feature} level (Line 3 in Alg.~\ref{alg:wenger}). Thus, it still permits a large number of FPs. 

\emph{How to reduce (and control) FPR?} 
One way to reduce the FPR in \cite{wenger2022data} is to employ a tighter significance level $\lambda$, which is used to detect whether the probability shift caused by the target feature is statistically higher than the non-target feature, and it defaults to be 0.1 in \cite{wenger2022data}.
However, this will also degrade the TPR, e.g., when we change to use a $\lambda$ of 1e-8, the FPR is reduced to 1.9\%\footnote{We also checked that larger $\lambda$ like 1e-10 cannot further reduce FPR.}, and the TPR is already degraded to $86.4\%$\footnote{Another option to reduce FPR is to adjust the proportion of positive tests ($\delta$) for the tool to return a positive outcome (details omitted): we increased $\delta$ from 0.6 (their default setting) to 1.0, and still observed a similar trade-off.}. 
In addition, while using a tighter significance level can reduce the FPR, it still \emph{cannot} control the FPR within a {specific} low regime (e.g., 0.01\%). 

To fulfill this requirement,  a principled way is to adapt a similar setup established in existing MI studies~\cite{ye2021enhanced,carlini2022membership,bertran2024scalable}. 
This procedure is sketched in Line 4 to Line 10 in Alg.~\ref{alg:wenger}. 
However, based on our results in the previous paragraph, it is likely that the TPR by their technique would further decrease when the TPR is controlled at a lower FPR regime (e.g., 0.01\%).

\textbf{Further comparison with \cite{wenger2022data}.} 
We further compare Wenger et al. with \sysname in our setting, where we evaluate how effective is their approach, if adapted for MI-based data provenance.

Specifically, we consider 100 target users and each user marks 25 samples (0.1\% of the training set). 
As in Wenger et al. \cite{wenger2022data}, we use ImageNet images as the outlier features to be blended into the CIFAR100 samples, and \sysname uses random color stripes as before. 
\cite{wenger2022data} have a configurable parameter to control the intensity of the outlier feature when overlaid to the target data - we thus compare \sysname with their approach under \textit{varying} marking intensities. 

\begin{figure}[t]
  \centering
  \includegraphics[ height=1.4in]{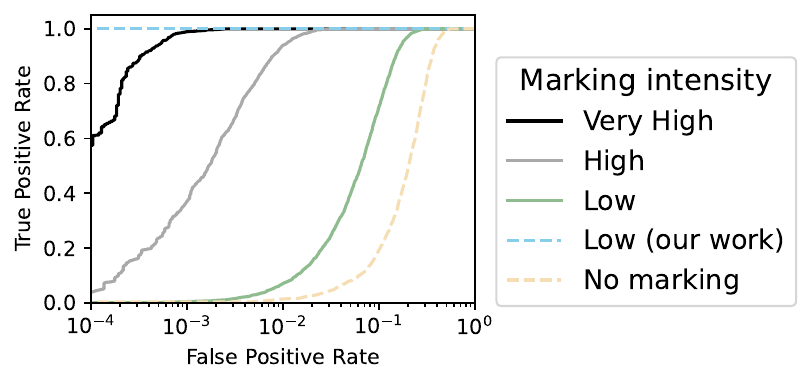}
  \caption{Comparing the technique by Wenger et al. (under varying data marking intensities) and \sysname in terms of their MI effectiveness for data auditing. \sysname achieves much higher TPR  than their approach in the low FPR regime.
  }
  \label{fig:wenger-mi}
\end{figure}

\begin{figure}[t]
  \centering
  \includegraphics[ height=1.5in]{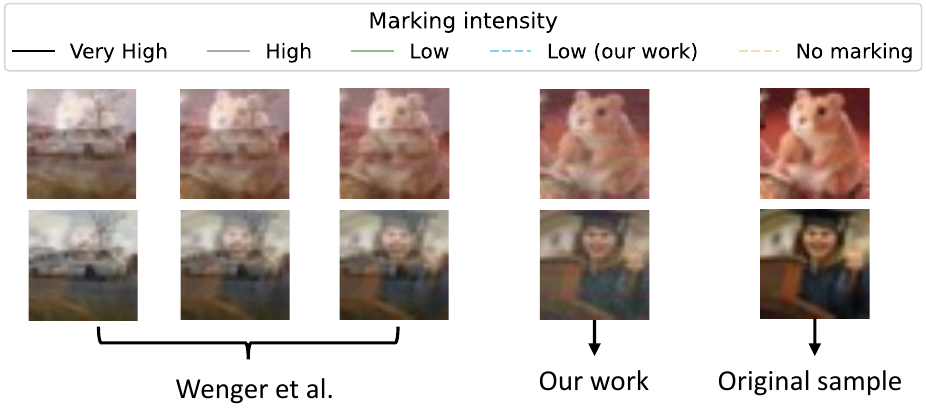}
  \caption{Samples marked with different intensities by the method of Wenger et al. vs. the samples marked by \sysname. 
  Our approach incurs significantly lower distortion to images than Wenger et al., and still outperforms their approach with much higher MI success. 
  }
  \label{fig:wenger-mi-exp}
    \vspace{-2mm}
\end{figure}

Fig.~\ref{fig:wenger-mi} shows the results, and Fig.~\ref{fig:wenger-mi-exp}. shows a visualization of the samples marked with different intensities.  

As shown in Fig.~\ref{fig:wenger-mi}, compared with the baseline approach without any data marking, the method by Wenger et al. can indeed  increase the MI success. 
For instance, the green line in Fig.~\ref{fig:wenger-mi} (marked using $m=0.6$, which is equivalent to the default marking intensity in \cite{wenger2022data}) increases the AUC from 0.7927 to 0.9253.
However, this still has very limited performance in the low FPR regime (e.g., 0\% TPR@0.01\% FPR).
Increasing the marking intensity in their approach can further improve the MI success, but it: 1) comes at the cost of higher image distortion; and 2) still has lower auditing performance than \sysname. 

For instance, under $m=0.5$ (meaning 50\% of the original features are replaced with the outlier features), their approach still yields much lower MI success than \sysname, e.g., 3.8\% TPR@0.01\% FPR vs. 100\% TPR by \sysname (the gray Vs. blue line in Fig.~\ref{fig:wenger-mi-exp}), and the samples marked by their approach already become highly conspicuous compared with  the original samples (the second and fifth column in  Fig.~\ref{fig:wenger-mi-exp}). Using higher marking intensity ($m=0.3$) still suffers from a similar trade-off.

Our earlier discussion in Section~\ref{sec:image-blending} explains that injecting outlier features alone is not enough, as the original features hinder the model's memorization on the outlier features. 
To overcome this,  we propose a novel approach to inject a small amount of \emph{perlin noise}, which works similar to adversarial perturbation and can significantly enhance the model's memorization on the marked samples (as evaluated earlier in Appendix~\ref{sec:ablation-two-step-marking}) while preserving high image quality (the fourth column in Fig.~\ref{fig:wenger-mi-exp}). 
Combined with the proposed set-based MI process (another key component in our work), \sysname is able to achieve perfect MI success (the blue line in Fig.~\ref{fig:wenger-mi}) and outperforms the method by Wenger et al.

\textbf{Summary.} 
While both Wenger et al. and our work aim at detecting the unauthorized data use in training DL models, their original technique greatly underestimates the detection false positives, and thus still struggles with achieving high TPR and low FPR. 

We also realign their technique for MI-based provenance. We find \sysname outperforms their approach by: 1) incurring lower distortion to the data; and 2) achieving significantly higher auditing performance. 
This renders \sysname a more practical data auditing tool.

\section{Comparison with Tramer et al.~\cite{tramer2022truth}} 
\label{sec:tramer-et-al-comparison} 
This section compares \sysname with another related work (Tramer et al). 
Their work proposes a technique to improve the MI success based on data poisoning. 
The main idea is to inject mislabeled samples into the model's training set, which can transform the target samples into outliers and amplify their influence to the model's decision, thereby making the target samples easier to be de-identified.  

Tramer et al. is the closest work to ours that can improve the MI success and without requiring the expensive shadow-model calibration. 
However, it assumes the users can mislabel their data, which can be challenging in the real-world scenarios where the users do not have control over the data labels~\cite{data-annotation-1,wenger2022data,deng2009imagenet}. 
In comparison, \sysname makes no such assumption and it only requires the users to mark their data with some small and targeted changes, which is a more realistic setting. 

For completeness, we  evaluate how effective their approach is, \emph{if} the users are able to mislabel their data for provenance purpose. 
We follow the targeted label flipping attack in Tramer et al., where for each target instance $(x,y)$, we inject multiple mislabeled data points $D_\text{adv}=\{(x,y'), ..., (x,y')\}$ for some label $y'\neq y$~\cite{tramer2022truth}. 
We consider 100 target users where each user injects 15 poisoned samples. We compare their approach with \sysname next. 

\begin{figure}[t]
  \centering
  \includegraphics[ height=1.5in]{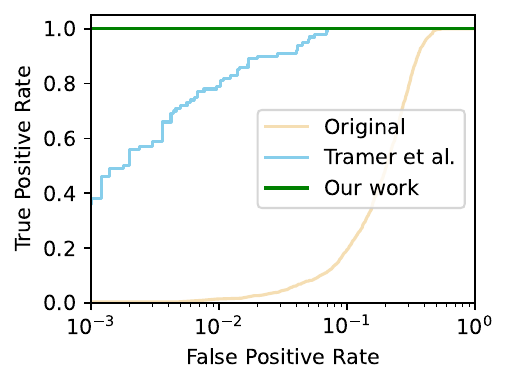}
  \caption{Comparing the performance between the approach by Tramer et al. and \sysname. \sysname achieves much higher TPR  than their approach in the low FPR regime. }
  \label{fig:tramer} 
\end{figure}

Fig.~\ref{fig:tramer} compares the performance by both approaches. 
As shown, the method by Tramer et al. indeed increases the exposure against the targeted samples. 
For example, under a 0.1\% FPR threshold, the TPR is increased from from 0\% to 36\%. 
Nevertheless, the increase of MI success for the method by Tramer et al. is still highly limited, while \sysname can achieve 100\% TPR@0.1\% FPR. 

Therefore, compared with the approach by Tramer et al., \sysname (1) represents a much more practical data provenance tool that does not require mislabeling any data, and (2) still achieves significantly higher MI success for tracing data provenance.

\section{{Evaluating Huang et al.~\cite{huang2024general} under Limited Data Setting}}
\label{sec:huang-et-al} 
Recent work by Huang et al. introduces a general framework for auditing unauthorized data use in training DL models~\cite{huang2024general}. Their main idea is to generate two perturbed copies of the target data, randomly publish one of them, and then  compare the model's membership score on the published Vs. the unpublished one. 

While both their work and \sysname have a similar goal, their work assumes the users can mark a large portion of data (1\%$\sim$10\%). However, in a real-world setting, model creators often curate their dataset by scraping data from multiple sources, and thus the data collected from each data holder may only constitute a small proportion of the  dataset. 
 For this, \sysname targets a more realistic (and challenging) scenario, where the users can modify only a limited amount of data ($\leq0.1\%$).

To understand the performance of their technique under such a limited-data setting, we use the original implementation from the authors~\cite{huang2024general} for an evaluation. 
{
Their approach can be applied to both image classifiers and foundation models, and we evaluate their approach  under the same classifier setting as \sysname (using CIFAR100). 
}
We first consider marking 1\% of the dataset, and find  that their technique can still maintain 100\% detection success rate. 

We then \emph{stress test} their technique using 0.1\% marking percentage, under which their auditing performance degrades considerably: the detection success rate is reduced to 20\% when the output vector is available. Under the label-only setting, the success rate further drops to 5\%. 
Therefore, our evaluation indicates that auditing data use under the limited-data setting still remains a major challenge, and \sysname represents a desirable technique that can offer superior auditing performance in these settings. 

With that said, their framework can still be used for other application domains that \sysname currently does not support, such as foundation models. In those applications, their tool remains the state-of-the-art data auditing solution, and we leave the extension of \sysname to other domains to future work.

\end{document}